\documentclass[12pt,preprint]{aastex}
 \usepackage{emulateapj5}
  \usepackage{apjfonts}
 \usepackage{epsf}

\shorttitle{Cosmic Ray Modified Shocks}
\shortauthors{Kang {\it et al.~}}
\slugcomment{To appear in ApJ Nov. 1, 2002}

\def\eg{{\it e.g.,}}
\def\ie{{\it i.e.,~}}

\def\cm3{~{\rm cm^{-3}}}

\def\lsim{\mathrel{  
        \raise0.3ex\hbox{$<$}\kern-0.75em{\lower0.65ex\hbox{$\sim$}}}}
\def\gsim{\mathrel{
        \raise0.3ex\hbox{$>$}\kern-0.75em{\lower0.65ex\hbox{$\sim$}}}}

\begin{document}
\title{Numerical Studies of Cosmic Ray Injection and Acceleration }

\author{Hyesung Kang}
\affil{Department of Earth Sciences, Pusan National University,
    Pusan 609-735, Korea} 
\email{kang@uju.es.pusan.ac.kr}

\author{T.W. Jones}
\affil{Department of Astronomy, University of Minnesota, Minneapolis, 
      MN 55455}
\email{twj@msi.umn.edu}

\and

\author{U.D.J. Gieseler}
\affil{Fachbereich Physik, Universit\"at Siegen, 57068 Siegen,  Germany}
\email{ug@nesa1.uni-siegen.de}

\altaffiltext{1}{Submitted to the Astrophysical Journal}

\begin{abstract}
A numerical scheme that incorporates a thermal leakage injection model 
into a combined gas dynamics and cosmic ray (CR, hereafter) 
diffusion-convection code has been developed.
The hydro/CR code can follow in a very cost-effective way
the evolution of CR modified planar quasi-parallel shocks by adopting subzone shock-tracking
and multi-level adaptive mesh refinement techniques.
An additional conservative quantity, $S= P_g/\rho^{\gamma_g-1}$, is
introduced to follow the adiabatic compression accurately in the 
precursor region, especially in front of strong, highly modified shocks.
The ``thermal leakage'' injection model is based on 
the nonlinear interactions of the suprathermal particles
with self-generated MHD waves in quasi-parallel shocks.
The particle injection is followed numerically by filtering 
the diffusive flux of suprathermal particles across the shock to the 
upstream region according to a velocity-dependent transparency function
that controls the fraction of leaking particles. This function
is determined by a single parameter, $\epsilon$, which should depend on 
the strength of postshock wave turbulence, but is modeled as a constant
parameter in our simulations.
We have studied CR injection and acceleration efficiencies during the
evolution of CR
modified planar shocks for a wide range of initial shock Mach numbers, $M_0$, 
assuming a Bohm-like diffusion coefficient.
For expected values of $\epsilon$ the
injection process is very efficient when the
subshock is strong, leading to fast and significant modification of the shock
structure.
As the CR pressure increases, the subshock weakens and the injection rate 
decreases accordingly, so that the subshock does not disappear.
Although some fraction of the particles injected early in the evolution continue to
be accelerated to ever higher energies, the postshock CR pressure reaches
an approximate time-asymptotic value due to a balance between fresh injection/acceleration and
advection/diffusion of the CR particles away from the shock. 
In the strong shock limit of $M_0 \gsim 30$,
the injection and acceleration processes are largely
independent of the initial shock Mach number for a given $\epsilon$,
while they are sensitively dependent on $M_0$ for $M_0<30$. 
We conclude 
that the injection rates in strong parallel shocks are
sufficient to lead to rapid nonlinear modifications to the
shock structures and
that self-consistent injection and time-dependent simulations
are crucial to understanding the non-linear evolution of CR modified
shocks. 
\end{abstract}


\section{Introduction}

Two decades ago it was recognized that 
CRs are probably produced very efficiently via diffusive shock acceleration 
(DSA) in ubiquitous astrophysical shocks, 
 \citep[for early reviews see, \eg][]{dru83,blaeic87,berzkry88}.
After the initial successes of the simple concept that
the particles can gain energy while temporarily trapped in the converging 
flows around a shock, it was quickly realized that
the full DSA treatment requires one to consider the complex nonlinear
interactions between energetic particles, resonantly scattering waves and
the underlying plasma \citep{madru01}.
One of the important aspects of those interactions
is the injection of suprathermal particles into the CR population at shocks.
According to quasi-linear theory as well as plasma simulations of strong 
quasi-parallel shocks, the streaming motion of superthermal particles against 
the background plasma can induce wave generation
leading to strong downstream MHD waves
that scatter particles and inhibit the particles from leaking upstream 
\citep[\eg][]{Bell78,Quest88}.
Particles in or close to the thermal population are especially restricted in this way.
As a consequence only a small fraction of suprathermal particles can swim upstream 
against the wave-particle interactions in the plasma flow and 
be injected into the higher energy CR population
to be further accelerated via the Fermi process.
The injection process and its efficiency control the amplitude of the CR population
and hence the degree of shock modification. 

Recently, significant progress in understanding this injection process 
in parallel shocks has been made through self-consistent, analytic, nonlinear calculations
by \citet{malvol95} and \citet{mal98}. 
The resulting theory has only one parameter; namely, the intensity
of the downstream waves, and that is tightly restricted, both by the
theory and by comparison with hybrid plasma simulations.
By adopting Malkov's analytic solution, we have developed a numerical treatment of
this injection model and incorporated it into a 
combined gas dynamics and CR diffusion-convection code \citep{gies00}.
According to the \citet{gies00} simulations, the injection process seemed to be 
self-regulated in such a way that the injection rate reaches and stays at 
a nearly stable value after quick initial adjustment, but well before the CR
shock reaches a steady state structure.
\citet{gies00} found about $10^{-3}$ of incoming thermal particles to be injected
into the CRs, roughly independent of Mach numbers.
However, due to severe computational requirements associated with the need to
resolve structures down to the physical shock thickness, those simulations
were carried out only until the characteristic maximum momentum of
$(p_{\rm max}/m_{\rm p} c)\sim 1$ was achieved.
Since strong shocks were still evolving at the end of the simulations,
the time-asymptotic limit could not be estimated for either the CR
acceleration efficiency or the CR spectrum.

Unlike ordinary gas shocks, the CR shock is a collisionless structure,
and includes a wide range of length
scales associated not only with the dissipation into ``thermal plasma'',
but also with the nonthermal particle diffusion process.
Those are characterized by the so-called diffusion lengths, 
$D_{\rm diff}(p) = \kappa(p)/u$, 
where $\kappa(p)$ is the spatial diffusion coefficient for CRs of momentum $p$, 
and $u$ is the characteristic flow velocity \citep{kanjon91}.
For strong scattering of suprathermal particles $D_{\rm diff}$ 
may not greatly exceed the physical dissipative, or ``gas'' shock thickness .
Accurate solutions to the CR diffusion-convection equation
require a computational grid spacing significantly smaller than $D_{\rm diff}$,
typically, $\Delta x \sim 0.1 D_{\rm diff}(p)$.
On the other hand, for a realistic diffusion transport model with a steeply
momentum-dependent diffusion coefficient,
the highest energy, relativistic particles have diffusion lengths
many orders of magnitude greater than those of the lowest energy particles. 

To follow the acceleration of highly relativistic CRs from suprathermal 
energies, all those scales need to be resolved numerically.
However, the diffusion and acceleration 
of the low energy particles are important only close to the shock owing 
to their small diffusion lengths. Elsewhere, they are effectively advected 
along with the underlying gas flow. 
At higher energies the needed resolution is less
severe, and on scales larger than $\sim D_{\rm diff}(p_{\rm max})$ 
the bulk flow and the nonthermal particles decouple, 
so resolution requirements are controlled by whatever factors
are necessary to define the broader flow properties.

Thus it is necessary to resolve numerically the diffusion length of the 
particles only around the shock. 
So, in \citet{kang01} we first implemented a {\it shock tracking scheme} 
\citep{levshy95} to locate the shock position exactly and then 
refine the grid resolution only around the shock
by applying multi-levels of refined grids \citep{berglev98}.  
The main properties of this code are:
1) the shock is tracked as an exact discontinuity,
2) a small region around the shock is refined with multi-level grids,
3) it is very cost-effective in terms of computational memory and time. 

In the present contribution, we have studied the 
CR injection and acceleration during the evolution of modified planar shocks
by implementing the numerical method 
for the thermal leakage injection model of \citet{gies00} into an
enhanced version of the CR/AMR 
hydrodynamics code, which we name CRASH (Cosmic Ray Amr SHock) code.
With our new CRASH code we were able to calculate the CR injection and 
acceleration efficiencies with a Bohm-like diffusion coefficient
for higher particle energies ($p/m_{\rm p}c \gg 1$) 
in shocks over a wide range of initial Mach numbers.
In the following section the basic equations solved in the simulations
are presented. 
We describe the thermal leakage injection model and its numerical
implementation in our CR transport code in \S 3.
We then outline our numerical methods in the CR/AMR hydrodynamics code in \S 4.
In \S 5 we present and discuss our simulation results, followed by 
a summary in \S 6.
Finally we discuss some test calculations in the Appendix.


\section{Basic Equations}
We solve the standard gasdynamic equations with CR pressure terms
added in the conservative, Eulerian
formulation for one dimensional plane-parallel geometry: 

\begin{equation}
{\partial \rho \over \partial t}  +  {\partial (u \rho) \over \partial x} = 0,
\label{masscon}
\end{equation}

\begin{equation}
{\partial (\rho u) \over \partial t}  +  {\partial (\rho u^2 + P_g + P_c) \over \partial x} = 0,
\label{mocon}
\end{equation}
 
\begin{equation}
{\partial (\rho e_g) \over \partial t}  +  {\partial (\rho e_g u + P_g u + P_c u) \over \partial x} = - L(x,t), 
\label{econ}
\end{equation}
where $P_{\rm g}$ and $P_{\rm c}$ are the gas and the CR pressure,
respectively, $e_{\rm g} = {P_{\rm g}}/{\rho}(\gamma_{\rm g}-1)+ u^2/2$
is the total energy density of the gas per unit mass and the rest of the 
variables have their usual meanings.
The injection energy loss term, $L(x,t)$, accounts for the
energy of the suprathermal particles injected to the CR component at
the subshock.
$L$ is nonzero only at the subshock transition, as explained
at the end of \S 3.2 and in \S 4.1.
CR inertia is neglected, as usual, in such computations.
A basic outline of the computational hydrodynamical scheme and its
adaptations for this problem are given in \S 4.1.

We do include here one unusual adaptation in the CRASH code that
significantly improves treatment of the CR precursor region.
In conventional conservative, gas dynamical numerical methods, the thermal energy is calculated by
subtracting the kinetic energy from the total gas energy $e_g$.
For highly supersonic flows where the kinetic energy is much larger than
the thermal energy, however, the numerical errors in calculating the kinetic gas 
energy could be much larger than the thermal energy itself.
Especially in strong CR modified shocks where the gas flow is compressed
through the precursor but remains highly supersonic, it is difficult to
follow this preshock compression accurately.  
In order to alleviate this difficulty, following the approach first 
introduced by \citet{rokc93}, we adopt an additional conservative quantity, 
the ``Modified Entropy'', 
\begin{equation}
S= P_g / \rho^{\gamma_g-1} , 
\end{equation}
related to the gas entropy per unit volume.
The modified entropy satisfies the following
conservation equation \citep{rokc93}:
\begin{equation}
{\partial S\over \partial t}  +  {\partial (uS) \over \partial x} =0,
\label{scon}
\end{equation}
which is valid outside the dissipative subshock; \ie where the gas entropy is conserved.
Hence the modified entropy equation is solved outside the subshock, while
the total energy equation is applied across the subshock. 
As we will discuss below, the injection process depends  
on the properties of the subshock and so on the precursor compression.
Our new hydro code with the modified entropy equation can accurately
follow the adiabatic compression through the precursor 
and provides a better method to follow the injection process.

The diffusion-convection equation for 
the pitch angle averaged CR distribution function $f(p,x,t)$ \citep[\eg][]{ski75} is given by
\begin{equation}
{\partial f\over \partial t}  + u {\partial f \over \partial x}
= {1\over3} ({\partial u \over \partial x})  p {\partial f\over
\partial p} +  {\partial \over \partial x} (\kappa(x,p)  {\partial f
\over \partial x}). 
\label{diffcon}
\end{equation}
and $\kappa(x,p)$ is the diffusion coefficient.
For convenience we always express the particle momentum, $p$ in
units $m_{\rm p}c$.
As in our previous studies, the function $g(p)=p^4f(p)$ is evolved instead
of $f(p)$ and $y = ln(p)$ is used instead of the momentum variable, $p$
for that step. This leads the equation solved to take the form
\citep{kanjon91, gies00}:
\begin{equation}
{\partial g\over \partial t}  + u {\partial g \over \partial x}
= {1\over3} ({\partial u \over \partial x})( {\partial g\over
\partial y} -4g) + {\partial \over \partial x} (\kappa(x,y)  {\partial g
\over \partial x}). 
\label{diffcon2}
\end{equation}
Our numerical approach to this equation is outlined in \S 4.1 with
additional details provided in the cited literature.

\section{Thermal Leakage Injection Model}
\subsection{Transparency Function}
In the ``thermal leakage'' injection model, most of the downstream thermal
protons are confined by nonlinear waves and only particles 
well into the tail of the Maxwellian distribution can 
leak upstream across the subshock.
Since any escaping particles have to swim against the advection 
flow downstream, the breadth of the thermal velocity distribution relative the
downstream flow velocity in the subshock rest-frame is central
to the injection problem.
In order to model this injection process in \citet{gies00} 
we adopted a ``transparency function'',
$\tau_{\rm esc}$, which expresses the probability that supra-thermal
particles at a given velocity can leak upstream through the magnetic waves,
based on non-linear particle interactions with self-generated waves 
\citep{mal98}.
In this scheme, the transparency function is {\it approximated} by
the following functional form,
\begin{eqnarray}
\tau_{\rm esc}(\epsilon,\upsilon/u_d)~=~H\left[ \tilde{\upsilon}-(1+\epsilon) \right]
        \left(1-\frac{u_d}{\upsilon}\right)^{-1}\,
        \left(1-\frac{1}{\tilde{\upsilon}}\right)\nonumber\\
 \cdot\, \exp\left\{-\left[\tilde{\upsilon}-(1+\epsilon)\right]^{-2}\right\}\,,
\label{transp}
\end{eqnarray}
which depends on the ratio of particle velocity, $\upsilon$, 
to downstream flow velocity in the subshock rest-frame, 
$u_d = u_s/r_{\rm sub}$, where $u_s$ and $r_{\rm sub}$ are the subshock velocity  
with respect to the immediate upstream flow and the subshock compression factor, respectively. 
Here $H$ is the Heaviside step function. 
The parameter, $\epsilon = B_0/B_{\perp}$, is defined
in \citet{mal98}, and measures the ratio of the amplitude of the 
postshock MHD wave
turbulence $B_{\perp}$ to the general magnetic field aligned with the
shock normal, $B_0$.
Here $\tilde{\upsilon}= \epsilon ~\upsilon/u_d$ is the normalized 
particle velocity. 

The lower right panel of Fig. 1 shows the Maxwellian distribution
function, $g_{\rm M}(p)=f_{\rm M}(p)p^4$ for the subshock Mach number 
$M_s=100$, and $\tau_{\rm esc}(p)$ as a function of $\upsilon/u_d$ 
for $\epsilon = 0.3$ (solid line) and $\epsilon=0.2$ (dashed line). 
The function $g_{\rm M}$ is plotted for $\beta=u_s/c=0.01$. 
The function $\tau_{\rm esc}$ plotted as a function of $\upsilon/u_d$
is independent of $\beta$.
The plot for $\tau_{\rm esc}(p)$ shows that the injection occurs 
predominantly for the particles with $ 5 u_d < \upsilon < 20 u_d$ where
$0< \tau_{\rm esc}(p) < 1$. 
The lower velocity particles with $\tau_{\rm esc} \approx 0$ represent
the gas particles, while the higher velocity particles with
$\tau_{\rm esc}\approx 1$ represent the CR particles.
So the particles are injected over a range of particle momenta
rather than a singly chosen injection momentum.  
Thus the transparency function eliminates the need to set up the lowest
momentum boundary that arbitrarily and abruptly distinguishes 
thermal plasma and accelerated particles.
If one takes a step function (\ie $\tau_{\rm esc}=0$ for $p<p_{\rm inj}$
and $\tau_{\rm esc}=1$ for $p\ge p_{\rm inj}$) rather than a smooth
function for the transparency function, the injection model becomes
the same as the earlier injection models where the particles are
injected at an injection momentum, $p_{\rm inj}$, which is then the momentum 
boundary between thermal and CR particles \citep[\eg][]{berz95, kanjon95}.

The only free parameter of the adopted transparency function is
$\epsilon$ and it is rather well constrained,
since $0.3\lsim \epsilon \lsim 0.4$ is indicated for strong shocks 
\citep{malvol98}.
Due to the exponential cut off in a thermal velocity distribution, however,
the injection rate depends sensitively on the value of $\epsilon$,
since the location in the Maxwellian tail distribution where 
$\partial\tau_{\rm esc}/\partial p \ne 0$ depends on $\epsilon$
(see the lower right panel of Fig. 1). 
Since the wave generation is relatively less efficient
for weak shocks, resulting in larger values of $\epsilon$,
this parameter should be dependent on the subshock Mach number.
However, since it is not certain precisely how $\epsilon$ should 
vary as the subshock becomes weaker with time,
we will keep the value of $\epsilon$ constant within each simulation
except for a few models where $\epsilon$ is changed during the simulation
to avoid a numerical problem during initialization.
We expect the injection rate
would be higher if we let $\epsilon$ increase as the subshock weakens,
compared with the calculation with a constant value. 
Although larger values of $\epsilon$ naturally lead to higher injection 
rate and higher acceleration efficiency, as we will show in \$ 5.1.2, 
the net enhancement of CR energy would be minimal (a few \%), 
since the CRs injected earlier are dominant
over the fresh particles injected recently at much weaker shocks
with a much smaller injection rate.
Another degree of freedom in this model is the detailed functional
form of the transparency function. The form given by
equation (\ref {transp}), which is an approximation of the
representation given in \citet{malvol98},
behaves like a smoothed step function which selects the range of
injection momenta for a given value of $\epsilon$.
With a modification in the shape of the transparency function
(due to a more accurate approximation),
we expect the resulting injection rate may not change significantly,
as long as the adopted $\tau_{\rm esc}$ has the characteristics of a
smoothed step function and picks the similar momentum range for the
injection pool.

\subsection{Numerical Scheme for Injection Model}
In \citet{gies00} we proposed a new numerical scheme for injection
to be used in kinetic equation simulations of diffusive shock
acceleration. The scheme emulates
thermal leakage injection through the transparency function given
by equation (\ref{transp}) for
particle leakage across a shock discontinuity. 
The total particle distribution function consists 
of Maxwellian and CR components ($g = g_{\rm M} + g_{\rm CR}$),
and there is no discrete momentum boundary between the two components.
In our scheme the Maxwellian distribution, $g_{\rm M}(p,x,t)$ is calculated from the 
density and temperature of the local gas at each time step
from gas dynamical data, since
it cannot be followed by the diffusion-convection equation (\ref{diffcon}). 
The suprathermal particles in this ``injection pool'' should have a 
significant anisotropic component, so the distribution there cannot be followed by
the diffusion-convection equation either.
In point of fact, one cannot cleanly
demark particles as thermalized and fluid-like or nonthermal and diffusive,
yet economical computational schemes exist only in those two limits.
That is the very reason we need an injection model as represented
here through the transparency function. 

We briefly explain how the distribution function is integrated in a way
that can implement the thermal leakage process in our CR transport code. 

1) The total particle distribution is updated in distinct ``advection and
diffusion'' steps for the momentum bins with  $\tau_{\rm esc}> 0$:
$g^{\rm n} \rightarrow \tilde g^{\rm n+1}$, where $g^{\rm n} $ is the value 
at the previous time step, and $\tilde g^{n+1}$ is an intermediate
value for $g^{n+1}$.
For thermal particles with $\tau_{\rm esc}=0$, 
$g^{\rm n+1} = g_{\rm M}^{\rm n+1}$.
Then the CR distribution up to this intermediate step is given by
$\tilde g_{\rm CR}^{\rm n+1} = \tilde g^{\rm n+1} - g_{\rm M}^{\rm n+1}$.

2) The net increase of CR particles, 
($\tilde g_{\rm CR}^{\rm n+1} -g_{\rm CR}^{n}$), calculated from
the diffusion-convection equation (\ref{diffcon}) should be reduced according
to the transparency function in the upstream region.
So an additional ``leakage'' step is applied to the CR distribution
function in the upstream region as  
\begin{equation}
g_{\rm CR}^{\rm n+1} = g_{\rm CR}^{\rm n} + \tau_{\rm esc}(t)\nonumber\\
   \cdot\left[\tilde g_{\rm CR}^{\rm n+1}
                         -g_{\rm CR}^{\rm n}\right].
\end{equation}
In other words, we first estimate the number of suprathermal particles 
that cross the shock according to the diffusion-convection equation, 
and then we allow only some fraction of the combined advective and diffusive 
fluxes to leak upstream with the probability prescribed by $\tau_{\rm esc}$. 
Obviously the CR particles with $\tau_{\rm esc}=1$ are not affected by
the leakage step.

We then need to estimate the gas energy loss rate at the subshock,
$L(x,t)$, due to particle leakage in order to conserve total energy. 
Although the injection process is realized numerically via the leakage 
of suprathermal particles to the upstream region (\ie diffusion in space),
we formulate the energy loss rate due to injection 
in terms of the particle acceleration in momentum space,
because leaking particles experience an adiabatic energy increase upon
crossing the subshock. 
The particle number flux crossing the momentum boundary at $p$ 
due to the adiabatic change of the momentum through the flow 
compression/expansion is given by $\tau_{\rm esc}(p)Q(p)$, 
where $Q(p) = -(4\pi/3)p^3f(p) (\partial u /\partial x)$,
since $\tau_{\rm esc}(p)$ gives the probability for the suprathermal
particles to cross the shock. 
Then the net change of the particle number density per unit time in
the momentum bin bounded by $p$ and $p+ {\rm d}p$ can be written as 
\begin{eqnarray}
 {\rm d}N_p = \tau_{\rm esc}(p+{\rm d}p)Q(p+{\rm d}p) -
                 \tau_{\rm esc}(p)Q(p)\nonumber\\
\approx \frac{\partial \tau_{\rm esc}}{\partial p}Q(p) {\rm d}p
     + \tau_{\rm esc}(p) \frac{\partial Q}{\partial p} {\rm d}p 
\label{dnp}
\end{eqnarray}
If the transparency function is a step function (\ie $\partial \tau_{\rm esc}
/ \partial p = \delta(p_{\rm inj})$ ),
the first term of equation (\ref{dnp}) gives the injection rate at $p=p_{\rm inj}$,
while the second term represents the acceleration of CRs for $p> p_{\rm inj}$.
For a smooth transparency function, the second term contains the
acceleration of both CRs and the suprathermal particles, 
since the momentum boundary between thermal and CR population is 
soft and $f(p) = f_{\rm CR}  + f_M$ for the momenta
in the injection pool (\ie $0< \tau_{\rm esc} (p) < 1$).
The injection rate due to the second term in equation (\ref{dnp})
that corresponds to the acceleration of suprathermal particles 
should be small because of exponential decrease of $f_M$ in this
momentum range.
As shown in Fig. 1, $\partial \tau_{\rm esc} / \partial p$, for  
the transparency function adopted here behaves almost like the delta function, 
and so the first term of equation (\ref{dnp}) dominates for the momenta 
in the injection pool.
So we take only the first term to estimate approximately the energy loss 
rate according to
\begin{equation}
L(x,t)\approx -\frac{2\pi}{3}~ m_{\rm p} c^2~ 
    \frac{\partial u}{\partial x}\int_{0}^{\infty}
     \frac{\partial \tau_{\rm esc}(p,t)}{\partial p} p^5 f(p,x,t)~{\rm d}p,
\label{einj}
\end{equation} 
where we have used the nonrelativistic approximation for the
kinetic energy of injected particles, $\frac{1}{2} m_pc^2 p^2$.
We remind the readers that $p$ is expressed in units of $m_{\rm p}c$.
The term $L(x,t) \Delta t$ is then subtracted from the gas thermal energy
in the immediate postshock region, and only there.
No matching term is necessary in updating the CRs, since $L$ is
introduced explicitly to account for evolution of $f(x,p,t)$ in
the ``injection pool'' where $f$ is in transition between
thermal and nonthermal forms.
Since the fraction of injected particles is very small, the energy 
exchanged through $L$ is, in any case, small, so that
it has little direct influence on the flow.

Since the diffusive flux of particles depends on the particle diffusion 
length, the CR injection rate numerically realized in our scheme 
depends on the ratio of $D_{\rm diff}(p_{\rm inj})/ \Delta x$ 
at the finest grid. However, the numerical results seem to be converged at
a spatial resolution of $\Delta x  \sim D_{\rm diff}(p_{\rm inj}) $.

\subsection{Dependence of Injection Rate on the Subshock Mach Number} 
As we will show below, the ensuing injection rate based on our thermal 
leakage model 
depends largely on the subshock Mach number, $M_s$, and
the inverse wave-amplitude parameter, $\epsilon$,
because the transparency filter is a function of 
$\tilde{\upsilon}= \epsilon ~\upsilon/u_d$.
Here we will explore the consequences of this dependence.
The thermal peak velocity of the Maxwellian distribution
of the immediate postshock gas relative to the mean flow is defined 
as $\upsilon_{\rm th}= 2 \sqrt {k_{\rm B} T_2 /m_{\rm p}}$,
where $T_2$ is the postshock temperature, $k_{\rm B}$ is the Boltzmann constant,
and $m_{\rm p}$ is the proton mass. The mean molecular weight is assumed to 
be one in this consideration and also in our numerical calculations.
The ratio of $\upsilon_{\rm th}/u_d$ is then determined only by 
$M_s$ as follows:
\begin{equation}
 {\upsilon_{\rm th} \over u_d} = \sqrt { \frac{12}{5}  {(5M_s^2-1)\over (M_s^2+3)}  }, 
\end{equation}
where $M_s$ is the subshock Mach number and we assume $\gamma_{\rm g} = 5/3$.
Since the injected particle flux is approximately proportional to 
$({\partial \tau_{\rm esc}}/{\partial p}) f(p) p^3{\rm d}p$ (see equation
[\ref{einj}]), we define the ``injection velocity'', $\upsilon_{\rm inj}$ as 
\begin{equation}
\upsilon_{\rm inj}= {
 {\int\limits_{0}^{\infty}
    \upsilon ~({\partial \tau_{\rm esc}}/{\partial p}) f_{\rm M}(p) p^3{\rm d}p}
 \over
 {\int\limits_{0}^{\infty}
     {\partial (\tau_{\rm esc}}/{\partial p}) f_{\rm M}(p) p^3 {\rm d}p} },
\end{equation}
where $f_{\rm M}$ is the Maxwellian velocity distribution of the immediate 
postshock gas.
One can easily show that the ratio of $\upsilon_{\rm inj}/u_d$ is a function of
$M_s$ and $\epsilon$ only (see Fig. 1).
Since in the model used here the rate of particle injection into the Fermi process is
proportional to the convolution of
${\partial \tau_{\rm esc}}/{\partial p}$ with the Maxwellian tail distribution,
the injection rate is determined mainly by the ratio of 
$\upsilon_{\rm inj}/ \upsilon_{\rm th}$,
rather than the velocities themselves.
This ratio depends on only $M_s$ and $\epsilon$.
We plot the values of $\upsilon_{\rm inj}/ \upsilon_{\rm th}$ in Fig. 1 for
$\epsilon = 0.2 - 0.35$. For strong shocks $M_s\gsim 10$ this ratio
becomes independent of $M_s$, while it increases for smaller $M_s$.
To illustrate, for $\epsilon=0.3$; $\upsilon_{\rm inj}/ \upsilon_{\rm th} = 1.87$
for $M_s\gsim 10$, while
$\upsilon_{\rm inj}/ \upsilon_{\rm th} \rightarrow 2.4$ as $M_s \rightarrow 2$.
Thus, the injection rate becomes independent of the subshock
Mach number for strong shocks. 
Larger values of $\upsilon_{\rm inj}/ \upsilon_{\rm th}$ for smaller $M_s$
means that the injection process is less efficient for weak shocks. 
This ratio is also larger for smaller $\epsilon$, since the leakage is prohibited
by stronger wave fields.
Thus, injection is more strongly inhibited by smaller $\epsilon$ (stronger 
waves).

Since the injected particles are nonrelativistic,
the ratio $\upsilon_{\rm inj}/ \upsilon_{\rm th}$, is in fact the same 
as the parameter $c_1= p_{\rm inj}/p_{\rm th}$ 
defined as a constant injection parameter in an earlier version of
a thermal leakage model by \citet{kanjon95} where 
values in the similar ranges (\ie $c_1=1.9-1.95$) were adopted.

As shown above, in a ``thermal leakage'' injection model where the leakage 
probability is determined by the ratio of the particle velocity to the 
advection velocity of the downstream flow relative to the subshock, 
we can expect the following :
a) the injection rate becomes independent of the subshock Mach number 
for $M_s\gsim 10$.
b) the injection is less efficient for weaker shocks ($M_s<10$) compared to 
stronger shocks.
c) in the time evolution of a CR modified shock
the injection rate decreases as the subshock weakens by way of precursor
compression.

\section{Numerical Methods and Model Parameters}
\subsection{CRASH: A CR/AMR Hydrodynamics Code with Modified Entropy}

Equations (1)-(3) and (5) are solved in two steps on an Eulerian grid: a gasdynamic
step without CR terms followed by a CR-dynamic step, as outlined
below.
Then equation (\ref{diffcon2}) is solved in two steps to update
the CR distribution and allow for injection at shocks. 
The advection term of equation (\ref{diffcon2})
is solved by the wave-propagation method, as for the
gasdynamic variables, except that only the entropy
wave applies. Then the diffusion term is solved by the semi-implicit
Crank-Nicholson scheme, as described in \citet{kanjon91}.

A full description of the combined gas dynamic/CR
algorithm without equation (\ref{scon}) can be found in \citet{kang01}.
We refer readers to that
work for most details and a discussion of various algorithm tests. 
Since we introduce two new features in the present work;
namely an improved treatment of the shock precursor and our new
``thermal leakage injection'' model, 
we briefly outline the basic scheme here. 
Some further, relevant code tests are described briefly in the Appendix.

In the gas dynamic step, the hydrodynamic conservation equations are solved 
by the explicit, LeVeque-Shyue Riemann-solver-based {\it wave-propagation} 
algorithm, temporarily ignoring the CR terms in eq. \ref{mocon} and \ref{econ}.
A nonlinear Riemann problem is solved at each interface 
between grid zones, and the wave solutions are used directly to update 
the mean dynamic variables in each zone. 
In regular zones not including a detected subshock
the gas pressure is updated from the modified entropy as $P_g = S~\rho^{
\gamma_g-1}$, where $S$ is updated by equation (\ref{scon}) outside the gas
shock (subshock). 
Updating the gas energy from equation (\ref{scon}) instead of equation (\ref{econ})
minimizes errors introduce to
the gas pressure due to dominant kinetic energy or $P_c$ gradients 
in the CR shock precursor.
For regular zones including a detected subshock, gas pressure is 
computed in the more conventional manner from the updated internal
energy found by equation (\ref{econ}). 
The latter is found by subtracting the kinetic energy 
estimated from equation (\ref{masscon}) and (\ref{mocon})
from the updated total energy (omitting
the $\partial(P_c u)/\partial x$ term across the subshock jump,
since by construction the diffusive CRs interact with the gas
only over distances exceeding the physical shock thickness).

The employed shock tracking method locates the exact position of the shock
within a regular grid zone using the shock speed obtained from the
nonlinear Riemann solver and then adds a new zone boundary there.
This subdivides a uniform zone of the underlying  regular grid into two sub-zones. 
In the next time step, this zone boundary (shock front) is moved to a new 
location using the Riemann solutions at the current shock location
and the waves are propagated onto the new set of grid zones \citep{levshy95}. 
Since the new grid locates the shock precisely
at an irregular zone boundary, 
the shock remains as an exact discontinuity without smearing.
The jump conditions across the discontinuous subshock are realized exactly,
because the wave solutions from the nonlinear Riemann solver 
(\ie speeds of waves and jumps associated
with three wave modes) are used directly to update the dynamic variables
in each zone.
Physically, the shock is not a discontinuity, of course, but a
relatively thin layer in which entropy is generated. Since CR shocks
are collisionless, this region is difficult to
define precisely. Nonetheless, a basic concept behind diffusive
shock acceleration theory is that we can identify this region and that
it is thinner than the interaction scale for even the lowest energy
CRs, given roughly by $D_{\rm diff}(p_{\rm inj})$. The need to
satisfy this constraint is the
main driver for incorporating the shock tracking scheme into the
CRASH code, since it minimizes the number of grid zones containing
the subshock.

In the CR-dynamic step, $\partial P_c / \partial x$ terms are added to the
gas momentum conservation equation (\ref{mocon}).  
The gas total energy, $e_g$ is updated by  
adding the new, CR-corrected, kinetic energy to the thermal energy.
At the subshock (only) the gas energy must also be reduced
according to the term $L$ in equation (\ref{einj})
to account for energy removed from the thermal population as
particles are converted into CRs (\ie as they are injected).
The quantity $L$ represents the energy per unit volume per
unit time transferred from gas to CRs in the immediate
postshock region. Physically this should occur over a distance
corresponding to the scattering lengths of injected particles,
or roughly $D_{\rm diff}(p_{\rm inj})$. In practice we spread this exchange
over three postshock zones of the finest refined grid (see the
following paragraph),
which is slightly broader, but reduces numerical difficulties
encountered if the energy is extracted too abruptly.
This step is analogous to the inclusion of radiative energy losses
in shocks, except that here the energy extraction is limited to
very close postshock distances.

In order to follow accurately the evolution of a CR modified shock,
it is necessary to resolve the precursor structure immediately
upstream of the subshock
and, at the same time, to solve correctly the diffusion of the low energy
particles near the injection pool.
Both of those have scales slightly greater than $D_{\rm diff}(p_{\rm inj})$.
While the full extent of the precursor increases with 
the diffusion length of the maximum accelerated CR momentum, 
the dominant scale length of the precursor as determined by
the gas pressure distribution is similar to an 
averaged diffusion length of the particle populations with the greatest 
contribution to the CR pressure.
As a result of these complications, a large dynamic range 
of resolved scales from the shock thickness to
the precursor scale is required for CR shock simulations.
To solve this problem generally, in order to allow eventually for
arbitrary and self-consistent diffusion models, we have successfully combined a
powerful ``Adaptive Mesh Refinement'' (AMR) technique
\citep{berglev98} and aforementioned ``shock tracking'' technique 
and implemented them into a hydro/CR code
based on the wave-propagation method \citep{kang01}.
The AMR technique allows us to ``zoom in'' inside the precursor structure
using a hierarchy of small, refined grid levels applied around the subshock.
The shock tracking technique follows hydrodynamical shocks within
regular zones and maintains them
as true discontinuities, thus allowing us to refine the region around
the gas subshock at an arbitrarily fine level.
The result is an enormous savings in both computational time and
data storage over what would be required to solve the problem using
more traditional methods on a single fine grid.

Fig. 2 shows the evolution of the gas density within the refinement region
on the $l_g= 1, 3, 5, 7$ level grids for a Mach 100 shock (see \S 4.4 for
detailed model parameters). There are 200 zones in each refined grid level
and $2\times10^4$ zones in the base grid.
The size of the refinement region is reduced by a factor of two between
adjacent grid levels, so the spatial resolution is refined by the same factor.  
Fig.2 demonstrates that the refinement region shrinks by a factor of $2^2$ for
$\Delta l_g=2$ and moves along with the shock.  
We also plotted the values at 200 zones with filled circles
at $\tilde t = 20$ in order to demonstrate that
the subshock is resolved as a discontinuous jump (see \S 4.3 for the
definition of time variable $\tilde t$).

\subsection{Bohm Diffusion Model}
The spatial diffusion coefficient can be expressed
in terms of a mean scattering length, $\lambda$, as
$\kappa(x,p) = {1\over 3} \lambda \upsilon$, where $\upsilon$ 
is the particle velocity.
The scattering length, $\lambda$, and thus $\kappa(x,p)$, should be
in principle determined by the intensity of resonantly interacting 
Alfv\'en waves.
For example, the Bohm diffusion model represents a saturated wave spectrum
and gives the minimum diffusion coefficient
as $\kappa_{\rm B} = 1/3 r_{\rm g} \upsilon$,
when the particles scatter within one gyration radius ($r_{\rm g}$)
due to completely random scatterings off the self-generated waves.
This gives
$\kappa_{\rm B} \propto ~{p^2}/{(p^2+1)^{1/2}}$.
We consider here only the proton CR component.
As stated before, we express particle momenta in units of $m_{\rm pc}$. 
In order to model amplification of self-generated turbulent waves 
due to compression of the perpendicular component of the magnetic field, 
the spatial dependence of the diffusion is modeled as
\begin{equation}
\kappa(x,p) = \kappa(p)({\rho_0 \over \rho(x)}),
\end{equation}
where $\rho_0$ is the upstream gas density. 
This form is also required to prevent the acoustic instability
of the precursor \citep{drufal86,kanjonryu92}.  

Here we define the acceleration time scale for a particle to reach momentum
$p$ \citep[\eg][]{dru83} as
\begin{equation}
\tau_{acc}(p)= {p\over<{dp/ dt}>} 
= {3\over {u_1-u_2}} ({\kappa_1\over u_1} + {\kappa_2\over u_2})
\approx {8\over u_s^2} \kappa_{\rm B}(p)~ ,
\end{equation}
where the subscripts, 1 and 2, designate the upstream and downstream
conditions, respectively.
The last expression is approximated for the velocity jump for a strong gas shock.  
For the Bohm diffusion coefficient the acceleration time of
relativistic particles increases linearly
with the momentum, so both the required simulation time and the
simulation box size 
increase linearly with the highest momentum to be accelerated. 
Hence it becomes extremely costly to extend the CR spectrum to $p_{\rm max} \gg 1$
even with our very efficient CRASH code.

\subsection{Model Parameters}

A CR modified shock has more complex structure than a discontinuous jump.
A precursor develops and the subshock weakens as the flow is modified 
by the CR pressure.
We will follow \citet{berzell99} to denote the values at the
far upstream as $u_0$ and $\rho_0$, the values immediately upstream of the
subshock as $u_1$ and $\rho_1$, and the values immediately downstream of
the subshock as $u_2$ and $\rho_2$.
We use $u_s$ as the subshock velocity relative to the immediate pre-subshock flow,
and $u_d$ as the immediate downstream flow velocity in the rest frame of the
subshock. 
The gas flow is compressed adiabatically in the precursor,
so the gas pressure immediately upstream of the subshock is given by 
$P_{\rm g,1} = P_{\rm g,0} (\rho_1/\rho_0)^{\gamma_{\rm g}}$.

The evolution of the CR modified shock depends on four parameters:
the gas adiabatic index, $\gamma_{\rm g}$; the
{\it initial} Mach number of the gas shock, $M_0=|u_0|/c_{\rm s,0}$;
$\beta=u_{\rm o}/c$; and the diffusion coefficient, $\kappa(p)$.
Here $u_0$ is the upstream flow velocity relative to the initial, 
unmodified shock, and $u_{\rm o}$ is the velocity normalization 
constant, which is set to be $|u_0|$. 
We assume $\gamma_{\rm g}=5/3$. 
For all simulations we present here $u_{\rm o}=$3000 kms$^{-1}$, so
$\beta=10^{-2}$.
We assume a Bohm-like diffusion coefficient,
$\kappa_{\rm B}=\kappa_{\rm o} {p^2}/{(p^2+1)^{1/2}}$.
The length and the time variables scale with $\kappa_{\rm o}$ through
the diffusion length and time
defined as $r_{\rm o} = \kappa_{\rm o}/u_{\rm o}$ and 
$t_{\rm o}=\kappa_{\rm o}/u_{\rm o}^2$, respectively.
Thus we do not need to choose a specific value for $\kappa_{\rm o}$,
as long as we adopt $r_{\rm o}$ and $t_{\rm o}$ as normalization constants
for length and time scales, respectively.

We consider a wide range of initial Mach numbers, 
$M_0= 2 - 200$ for the {\it initial} shock jump 
by adjusting the initial preshock gas pressure by
\begin{equation}
P_{\rm g,0}= 0.75/(1.25 M_0^2 -0.25).
\end{equation}
The far upstream density and flow velocity are, in code units, fixed at 
$\rho_0=1,$ $u_0= -1$.
Thus the preshock gas is colder for stronger shocks,
while the gas density and the shock velocity are the same
for all models. 
Then the initial jump conditions for downstream region in the rest frame of 
the shock are given as:
$\rho_2=r$, $u_2= -1/r$, $P_{\rm g,2}= 0.75$ in downstream region,
where the compression ratio is $r= 4 M_0^2 /(M_0^2 + 3)$.
Since the subshock velocity changes as the subshock
weakens in response to growing upstream CR pressure,
the subshock drifts in the initial shock rest frame.
In order to keep the subshock close to the center of the simulation box
a small velocity shift ($u_{\rm shift}$), which is experimentally determined, is 
added occasionally to the rest frame of the initial shock. 
There are no pre-existing CRs in the simulated flows, 
so the CRs are introduced to the system 
at the shock via thermal leakage as the system evolves.

Although the theoretically preferred values of $\epsilon$ lie between 
0.3 and 0.4 for strong shocks \citep{mal98}, 
such values lead to very efficient initial injection for shocks of $M_0 \gsim 100$.
Efficient injection at the start of the simulation results in the 
rapid surges of $P_c$ and the precursor 
at the early stage, which, in fact, are too fast for the code to follow accurately.
During this early development the gradients in the precursor are extremely steep.
Consequently, the flow structure on the base grid, which is calculated from the average
of $2^{l_{\rm max}}$ zones of the finest grid, develops large discrete 
jumps. The Riemann solver for such flows can fail to converge. 
For $\epsilon=0.25-0.3$ we can use up to $l_{\rm max}=5$ levels of refined 
grids for $M_0 \le 100$ shocks without failures of the Riemann solver. 
For $\epsilon=0.17-0.2$, $l_{\rm max}=7$ can be used for $M_0 \le 200$ shocks.
So we consider two sets of models depending on the injection rate, that is,
models with $\epsilon=0.25-0.3$ and models with $\epsilon=0.17-0.2$.
While these parameter values were applied for practical reasons, the
thermal leakage is expected to be more difficult in oblique shocks, so
these runs can offer some insights into the importance of that
effect.

For the injection models with $\epsilon=0.25-0.3$,
the simulations were carried out on a base grid with
$\Delta x_0 = 3.2\times 10^{-3}$ using $l_{\rm max}=5$ additional grid levels,
so $\Delta x_5 = 10^{-4}$ on the finest grid.
Here spatial resolutions are given in units of $r_{\rm o}$. 
The simulated space is $x=[-12.8,12.8]$ and $N=8000$ zones are used 
on the base grid.
We have repeated one of the models ($M_0=30$ case) with $l_{\rm max}=4$
and found that the numerical results are well converged within 1 \%
in terms of CR pressure at this spatial resolution. 
The number of refined zones around the shock is $N_{rf}=200$ on the base
grid and so there are $2N_{rf}=400$ zones on each refined level.
The length of the refined region at the base grid is 0.64, so $1/40$ of the
entire simulated space on the base grid is refined. 
In these cases we integrate the simulation until $\tilde t\equiv t/t_{\rm o}=20$, so that the maximum
momentum achieved by the end of simulation is of order of $p_{\rm max}
\sim 4$, above which the CR distribution function decreases exponentially.

For the injection models with $\epsilon=0.17-0.2$,
the base grid spacing is $\Delta x_0 = 1.28\times 10^{-2}$
with $l_{\rm max}=7$, so $\Delta x_7 = 10^{-4}$ on the finest grid.
The simulated space is $x=[-128.,128.]$ and $N=20000$ zones are used 
at the base grid.
For these models $N_{rf}=100$ on the base
grid and $2N_{rf}=200$ at each refined level.
The length of the refined region at the base grid is 1.28, so only $1/200$ 
of the entire simulated space on the base grid is refined. 
We integrate these injection models until $\tilde t=100-300$. 
For all models we use 230 logarithmic momentum zones 
in the interval $\log (p/m_p c)=[\log p_0,\log p_1]=[-3.0,+3.0]$. 

Gasdynamic variables are assumed to be continuous at both left and
right boundaries of the simulation box.
For the CR distribution function a continuous boundary is assumed for 
the advection step and a no-flux boundary condition is adopted
for the spatial diffusion step.  
Either below $p_0$ or above $p_1$;
$g(p)=0$ is assumed. 
For clarity we distinguish here between the highest momentum
included in the simulations, $p_1$ and the value of the momentum
$p_{\rm max}$ characterizing the effective upper cutoff in the
actual distribution, $g(p)$. In practical terms
there generally is a clear momentum above which $g(p)$ drops
sharply, so that we identify as $p_{\rm max}$.
The maximum momentum $p_{\rm max}$
will generally evolve towards, $p_1$, but in all the 
simulations reported here remains well 
below it at the terminal time. 

\subsection{Injection and Acceleration Efficiencies}

The fraction of particles that have been swept through the shock after the
time $t$, and then injected into the CR distribution is estimated by
\begin{equation}
\xi(t)=\frac {\int {\rm d x} \int 4\pi f_{\rm CR}(p,x,t)p^2 {\rm d p}}
{ \int n_0 u_0^\prime (t^{\prime}) {\rm d t^{\prime}}}\,
\end{equation}
where $f_{\rm CR}$ is the CR distribution function, $n_0$ is the
particle number density, and $u_0^\prime (t)$
is the instantaneous shock velocity relative to the far upstream flow. 
As the CR pressure becomes dominant for strongly CR mediated shocks,
the effective adiabatic index of the flow becomes smaller than 5/3.
This combines with other factors discussed below that allow the
total shock compression to increase and
causes the subshock to slow down with respect to the upstream gas. 
So the instantaneous shock velocity decreases as the CR pressure increases. 
One should also note that $\xi(t)$ represents 
a ``moving time-averaged'' injection efficiency rather than a current
instantaneous value.

As a measure of acceleration efficiency,
we define the  ``CR energy ratio''; namely the
ratio of the  total CR energy within the simulation box to 
the kinetic energy in the {\it initial shock rest frame} 
that has entered the simulation box from far upstream, 
\begin{equation}
\Phi(t)=\frac {\int {\rm d x} E_{\rm CR}(x,t)}
{ 0.5\rho_0 |u_0|^3 t }\,.
\label{crenrat}
\end{equation}
The total energy in the simulation box is conserved in the
initial shock rest frame, since the total energy flux entering the simulation
from far upstream, 
$0.5\rho_0u_0^3 + \gamma_{\rm g} P_{\rm g,0}u_0/(\gamma_{\rm g}-1)$, 
is balanced by the same flux exiting the grid far downstream.
We intentionally placed the boundaries well away from the shock 
to avoid boundary issues.
Since our shock models have the same upstream density and velocity, but
different gas pressure depending on $M_0$, 
we use the kinetic energy flux rather than the total
energy flux to normalize the ``CR energy ratio''.
For strong shocks the total energy flux becomes the same as 
the kinetic energy flux.

Since the shock slows down due to nonlinear modification, 
the kinetic energy flux in the instantaneous
shock rest frame also decreases.  
So we also calculate the ratio of the total CR energy in the simulation box 
to the kinetic energy defined in the {\it instantaneous 
shock frame} which has entered the shock from far upstream, 
\begin{equation}
\Phi^\prime (t)=\frac {\int {\rm d x} E_{\rm CR}(x,t)}{ 
 \int 0.5\rho_0 u_0^\prime (t)^3 {\rm dt}}.
\end{equation}

\section{Simulation Results}

\subsection{Models with $\epsilon=0.25-0.3$}

\subsubsection{Shock Structure Modification} 
Figs. 3 and 4 show time evolution of the CR modified shock structure 
for $M_0=5$ and $M_0=30$ shocks, respectively, with $\epsilon=0.3$. 
As CRs are injected and accelerated, the CR pressure increases, 
the precursor grows, and the subshock slows down and weakens.  
Nonlinear modification to the shock structure is much stronger
and the subshock drift is greater for higher Mach number models.  
At the simulation ends the subshock drifts to the left 
with $u_{\rm drift}\approx -0.14$ for the $M_0=30$ shock
and with $u_{\rm drift}\approx -0.038$ for the $M_0=5$ shock.
Since the initial simulation frame was set up with the $u_{\rm shift}=0.01$ 
relative to the initial shock rest frame, the subshock drifts in the initial
shock rest frame with $u=u_{\rm drift}-0.01$.
Thus for strong shocks
the subshock velocity relative to far upstream flow
becomes $u_0'\sim 0.85$ instead of $|u_0|=1$, approximately
independent of $M_0$.
For low Mach number shocks, however, this velocity difference is smaller and 
increases with $M_0$ (see also Fig. 12 below).

Due to strong injection, the CR pressure increases and the modification
to the flow structure proceeds very quickly, before $\tilde t \lsim 0.1$.
After the initial quick adjustment, the CR pressure at the shock reaches 
{\it approximate} time-asymptotic values when the fresh injection
and acceleration are balanced with advection and spreading of high energy particles 
due to strong diffusion.
This balance seems to occur earlier in time for larger values of $\epsilon$ 
for models of a given Mach number (see also Fig. 5 below).
For a given value of $\epsilon$, it occurs gradually earlier 
with increasing initial Mach number, but at a similar time in the strong shock limit. 
We will show below that the time-asymptotic values of $P_c$ are related to the 
subshock evolution, as well. 
After $P_{\rm c,2}$ becomes {\it quasi-steady},
the overall shock structure evolves approximately in a {\it self-similar} way,
as shown in Figs. 3-4.
This behavior can be understood 
from the fact that $P_{\rm c,2}$ at the shock stays constant 
and the characteristic scale length of $P_c$ and precursor scale length increase
linearly with time, since the diffusion length of the highest momentum, 
$D_{\rm diff} (p_{\rm max}) \propto t$ for $p_{\rm max} \gg 1$ 
and the advection length also scales with time.

The evolution of the density distribution in Fig. 4 includes the formation of
the postshock ``density spike'' noted in several previous time dependent
studies \citep[\eg][]{jk90}. The density spike forms because the 
weakening of the subshock lags behind formation of the precursor
during initial shock modification. For strong shocks this can
produce a substantial, temporary ``overshoot'' in the total compression (see also Fig. 9).
That feature eventually detaches from the steady shock and is carried 
downstream from it.

Evolution of subshock and precursor properties can be seen in Fig. 5, 
where the subshock velocity, $u_s$ and  Mach number, $M_s$, as well as
the pre-subshock gas density,
$\rho_1/\rho_0$, and the immediate postshock gas density, $\rho_2/\rho_0$, 
are plotted against time for the models of $M_0=2-30$
and $\epsilon=0.3$. 
As mentioned before, the modification to the flow structure occurs 
mostly before $\tilde t \lsim 0.1$, and so
the subshock velocity decreases and the precursor compression increases
at the same time.
After $\tilde t > 0.1$, 
the subshock Mach number remains nearly constant at $M_s\sim 2-3$ for these shock models,
and the subshock persists. Thus, a completely smooth transition never develops.
The ratio $\rho_2/\rho_0$ is greater for
higher Mach number shocks and in fact follows roughly $M_0^{0.6}$
scaling relation (see \S5.3 for further discussion).
In our simulations, the compression continues to grow slowly with time,
especially for strong shocks, even after
the overall shock structure measured from far upstream through the
subshock appears mostly to have reached a ``quasi-steady state''.

Obviously the compression factor is much higher than what is expected in a 
{\it steady-state}, energy conserving shock jump for relativistic gas 
(\ie $\rho_2/\rho_0= 7$ with
$\gamma_c=4/3$). Similar results to ours have been found previously in
spherically expanding CR dominated shocks \citep{berz95} and in
steady state planar shocks with strongly momentum dependent
diffusion when provision is made for energy losses carried
by escaping high momentum particles \citep{mal97, berzell99, madru01}. 
While our shocks are not
yet truly steady, they have all evolved well past the transient
stage associated with the density spike mentioned above. 
It is very unlikely that the density jumps across our
simulated shocks would decrease were they allowed to evolve over
much longer times. As noted, compression appears still to be increasing.
On the
other hand, the spatial and momentum boundaries in our
simulations are such that energy losses for the entire
system can be ignored, so they might be
considered to be conservative. 

The key feature in common to 
the CR shock solutions with non-transient large density jumps is
that they all involve significant energy transport away from the transition
itself. \citet{dvb95} pointed out that subshock smoothing in
spherical systems was inhibited by the fact that CRs
diffuse into an ever-increasing volume, thus extracting
energy from the shock itself. Steady planar shocks with 
either a spatial or a momentum ``escape'' boundary that
allows significant energy to leave the shock structure
encounter the same effect. Once the momentum
distributions in our CR populations near the shock
evolved so that $p_{\rm max} \sim p_1$ they would
be influenced by similar ``escape'' losses. In the meantime
they appear to be evolving towards properties
that are consistent with those losses. At the
``intermediate ages'' we see in our shock
evolutionary histories, energy is not
leaving the system, but is being transported 
out of the region of the subshock by diffusion of
the highest energy CRs. The quasi-steady structure
is growing in a self-similar manner, in fact, 
reflecting the linear increase in $D_{\rm diff} (p_{\rm max})$
with time. 
Since advection extends the downstream region in a similar way,
the immediate postshock properties can remain essentially constant even while the
global structure continues to grow in phase space.

Regardless of the quasi-steadiness of the postshock CR pressure and the self-
similarity of the flow structure, 
the CR distribution function continues to become harder and
extends to ever higher momenta until scattering
becomes too weak to contain the particles.  
Thus, the adiabatic index of the CRs, $\gamma_c = 1+P_c/E_c$, decreases with time at the shock
and tends towards $4/3$ as the CR pressure and energy become dominated
by relativistic particles.
This evolution in $\gamma_c$ is more dramatic away from the subshock and
especially upstream, since spatial diffusion is biased towards more
energetic particles.
This secular evolution of $\gamma_c$ affects the precursor compression and
causes the subshock properties to evolve slowly, especially for strong shocks,
as can be seen in Fig. 5.
Although we adopted the modified entropy equation to follow the precursor
adiabatic compression accurately,
the subshock properties are not numerically steady but somewhat noisy.
This unsteady behavior of the subshock seems to be 
due to the CR pressure gradient, $\partial P_c/ \partial x$,
which is separately added to the momentum conservation equation.
Within the precursor this pressure force is large and compresses
the gas adiabatically, so it determines the position and
the properties of the subshock.
Thus, inside the precursor the accuracy in calculation of $P_c$ 
distribution would control the subshock calculation 
rather than the accuracy in calculation of the gas adiabatic 
compression. 
If one used a conventional code with the total energy equation alone,
these quantities would be much noisier than what is seen in Fig. 5.

\subsubsection{Injection and Acceleration Efficiencies} 

Fig. 6 shows how the CR energy ratio, $\Phi$, 
the CR pressure at the shock normalized to the far upstream ram pressure 
in the instantaneous shock frame, $P_{\rm c,2}/\rho_0 u_0^\prime (t)^2$,
and the ``time-averaged'' injection efficiencies, $\xi$,
evolve for shocks with different Mach numbers when $\epsilon=0.3$ (left three panels)
or $\epsilon=0.25$ (right three panels) is adopted.
As mentioned in the previous section,
for all Mach numbers the postshock $P_c$ increases until
a balance between injection/acceleration and advection/diffusion of CRs
is achieved, and then stays at a steady value afterwards.
The time-asymptotic value of the CR pressure becomes, 
$P_{\rm c,2}/\rho_0 u_0^\prime (t)^2 \sim 0.8$,
for $M_0= 30$ with $\epsilon=0.3$.
As seen in the previous section, the subshock slows down, 
and the subshock velocity relative to the far upstream flow 
becomes $u_0^\prime \sim 0.85 |u_0|$ for $M_0=30$. 
Since the subshock has reached a ``quasi-steady state'' in our simulations,
we can apply the mass and momentum flux conservation in the subshock rest frame as
\begin{equation}
  \rho_0 u_0' = \rho_1 u_s = \rho_2 u_d ,
\end{equation} 
\begin{equation}
 \rho_2 u_d^2 + P_{\rm g,2} + P_{\rm c,2} = \rho_0 u_0'^2 + P_{\rm g,0}.
\end{equation} 
The postshock gas pressure is given by the jump condition,
$P_{\rm g,2}= (1.25M_s^2-0.25)P_{\rm g,1}$, 
so it can be approximated as $P_{\rm g,2} \approx 0.75 \rho_1 u_s^2 $ for 
$2<M_s<4$ for models considered here (see Fig. 5).
Hence for large $M_0$ the postshock CR pressure can be approximated as
$ {P_{\rm c,2}/{\rho_0 u_0'^2}} \approx  1 - 0.75(u_s/u_0') - (u_d/u_0')$.
Since $u_0'\approx 0.85|u_0|$ and $u_s\approx 0.17 |u_0|$ for $M_0=30$ shock, 
$P_{\rm c,2}/\rho_0 u_0'^2 \sim 0.8$, which is consistent with what
we found in our simulations.

The CR energy ratio, $\Phi$, increases with time, but asymptotes to
a constant value, once $P_{\rm c,2}$ has reached a  quasi-steady value.
As mentioned in the previous section,
this results from the ``self-similar'' behavior of the $P_{\rm c}$ 
distribution. 
Since both the numerator and denominator in equation (\ref{crenrat}) increase 
approximately linearly with time in the case of the self-similar evolution 
of $E_{\rm c}$, $\Phi$ becomes steady when $P_{\rm c,2}$ becomes constant.
Time-asymptotic values of $\Phi$ increase with the initial Mach number $M_0$ and
$\Phi\approx 0.53$ for $ M_0=30$ at the terminal time (see also Fig. 12).
On the other hand, time asymptotic values of $\Phi^\prime$ become
$\Phi^\prime \approx \Phi \cdot (u_0^\prime/|u_0|)^3 \sim 0.86$ 
for the $M_0=30$ shock model. 

The initial injection rate before the weakening of the 
subshock is roughly independent of $M_0$ for strong shocks, 
as we expect from the discussion in \S 3.3. 
Initially it briefly increases to $\xi \sim 0.01$  for strong shocks until
$ t\sim \tau_{\rm acc}(p_{\rm inj})$, the acceleration time scale of the 
injection momenta.
The initial brief rise in $\xi$ is a numerical artifact occurring in the 
interval when the initially pure Maxwellian distribution adjusts to form a
nonthermal tail in the momentum range where injection takes place.
Afterwards it decreases as the subshock weakens, since the injection
is less efficient at low Mach number shocks.
After the initial adjustment period
the injection rate decreases as a power-law form with time,
$\xi \sim \xi_o \tilde t^{-\alpha}$ for $0.05<\tilde t < 20$,
where $0.15\lsim\alpha\lsim 0.4$ with larger values for higher $M_0$.
Although the subshock Mach number decreases to similar values
after $\tilde t =0.1$ for all models (see Fig. 5),
the injection rate decreases faster for higher $M_0$ models.
This can be explained as follows.
The injection momentum $p_{\rm inj}$ after $\tilde t=0.1$ is smaller
for higher $M_0$, since the subshock velocity is reduced further
in response to stronger structure modification.
Diffusive particle flux upstream, which is realized numerically in our code
with a given subshock thickness ($\Delta x_5$), is reduced if
$D_{\rm diff}(p_{\rm inj}) < \Delta x_5$. 
The resulting injection rate is smaller for higher $M_0$ models
in which the injection momentum is lower.
Thus the strong decrease of $\xi$ for high $M_0$ models below the level of 
$\xi$ for $M_0=3$ model should be regarded as a numerical artifact due
to a fixed shock thickness.

In order to explore the dependence of our injection model on the parameter
$\epsilon$, we also considered the same shock models with $\epsilon=0.25$
and show $\Phi$, $P_{\rm c,2}$ and $\xi$ in the right three panels of Fig. 6.
Because of less efficient injection compared to $\epsilon=0.3$ cases, 
the initial increases of $P_{\rm c,2}$ and $\Phi$ are delayed somewhat and 
their time-asymptotic values are reduced a little, as well. 
Before $\tilde t \sim 1$, the CR pressure is dominated by 
non-relativistic particles and the details of the injection history
are important to the shock structures.
Afterwards, the relativistic particles dominate and
the results become less sensitive to the values of $\epsilon$ and 
the details of the injection process. 
For example, the injection rate $\xi$ is 2-3 times lower with $\epsilon=0.25$ 
than with $\epsilon=0.3$ for the $M_0=30$ shock,
but the time-asymptotic value of $\Phi$ is smaller only by 2\%.
In summary, the CR energy ratio, $\Phi$, depends sensitively 
on $M_0$ for weak shocks, then becomes independent on $M_0$ for
strong shocks, but depends only weakly on the injection rate ($\epsilon$).

\subsubsection{The Particle Distribution} 
Time evolution of the distribution function $g(p)=f(p)p^4$ {\it at the shock}
is shown in Fig. 7 ($M_0=5$) and Fig. 8 ($M_0=30$)
along with the changes in the transparency function $\tau_{\rm esc}$. 
The Maxwellian distribution shifts to lower momenta primarily due to
weakening of the gas subshock in response to flow
deceleration in the precursor, but also due to the loss of thermal energy
to the injection process.
As the subshock weakens the postshock flow speed in the subshock rest
frame, $u_d$, decreases, so the transparency function also shifts to lower
momenta. 
As shown in Fig. 1, the ratio of $\upsilon_{\rm inj} / \upsilon_{\rm th}$
increases as the subshock weakens, leading to decreases in the 
injection rate.
For momenta just above the injection pool, the distribution function changes smoothly
from a Maxwellian distribution to an approximate power-law whose index is
close to the test-particle slope for the subshock.
However, we note that the particle distribution at lower energy 
does not evolve as a simple power-law, but, instead, it depends on 
the time-dependent injection history, since the subshock properties
and $\tau_{\rm esc}(u_d)$ vary with time and their numerical solutions
in our simulations contain low-amplitude noise. 
Those ``historical details'' become smeared out at higher momenta.
While a fraction of particles injected earlier continues to be 
accelerated to higher momenta, so that $p_{\rm max}(t)$ increases, 
the amplitude of $g(p)$ at the shock and at a given momentum 
decreases with time for $p<p_{\rm max}(t)$ due to diffusion.

These results can be compared with those of \citet{gies00} in which
a hydrodynamic code based on the Total Variation Diminishing (TVD) scheme
was used. 
In \citet{gies00} the diffusion coefficient was 100 times larger than the value
used in the present study, so in a comparison
the length and time scales should be scaled
by that amount. 
The main difference between the current CR/AMR code and the CR/TVD code
is that the subshock is maintained as a discontinuous jump in the CR/AMR code.
Numerical gradients defined at the subshock, therefore, span a single
regular zone.
Given the same nominal spatial resolution, the shock is more sharply resolved
in the CR/AMR simulations.  This affects the diffusion of suprathermal
particles near the shock and so leads to higher injection rate.

\subsection{Models with $\epsilon=0.17-0.2$}
In order to extend the calculation to much longer simulation times,
so that much more energetic particles will be accelerated,
we need to include a larger computational domain because of the fast
spatial diffusion of the highest energy particles.
Thus, the computational time for such calculations becomes very demanding. 
We increase the refinement level to $l_{\rm max}=7$ and 
consider a computation domain 10 times larger than that of the stronger
injection calculations discussed in \S5.1.
This forces us to choose smaller values of $\epsilon=0.17-0.2$ 
(weaker injection rate) to avoid the initialization
numerical problem mentioned in \S 4.3. 
For $M_0= 5-100$ $\epsilon=0.2$ is used, while 
for $M_0$=150 and 200 $\epsilon=0.17$ is
used for $\tilde t \le 1$ then increased to $\epsilon=0.2$ for $\tilde t >1$.

\subsubsection{Shock Structure Modification} 
Fig. 9 shows the evolution of the shock structure for a $M_0=200$ shock
at much later times up to $\tilde t= 140$.
Overall shock structures reach {\it approximate} time-asymptotic states 
after the initial quick adjustment, 
and then follow ``self-similar'' evolutions afterwards as in the 
the $M_0=30$ shock with $\epsilon=0.3$ shown in Fig. 4.
The subshock slows down and drifts with the velocity 
$u_{\rm drift}\approx -0.025$ after the shock structure evolution became
quasi-steady for the strong shock models
and the starting velocity shift relative to the initial shock rest 
frame was $u_{\rm shift}= 0.145$. 
So the time-asymptotic velocity of the subshock relative to far upstream 
becomes $u_0' \approx 0.83$ for $\epsilon=0.2$ for strong shocks (see also
Fig. 12).
This shock speed was $u_0'\approx 0.85$ for $M_0=30$ shock model with 
$\epsilon=0.3$.
The subshock Mach number decreases with time, but reaches 
quasi-steady values after $\tilde t =100$.
The values of $M_s$ at $\tilde t=100$ seem to scale with the initial shock
Mach number $M_0$ roughly as $M_s \sim 2.9 M_0^{0.13}$.
Although the postshock CR pressure and the subshock speed become roughly
quasi-steady after the initial fast modification stage, 
the precursor compression continues to increase up to the terminal time
and the rate of this increase is higher for higher $M_0$ models.
We see that $\gamma_c$ is the largest near the shock where the injected
suprathermal particles are dominant. But it becomes smaller away from the
shock, because relativistic particles dominates there.
The spatial distribution of $P_c$ peaks sharply from upstream at the subshock 
on account of the small
diffusion length scale reflecting the dominance of low energy
particles near the subshock.  

\subsubsection{Injection and Acceleration Efficiencies} 
The evolution of $\Phi, P_{\rm c,2}$, and $\xi$
for $M_0=5-200$ with $\epsilon=0.2$ is shown in the left three panels of Fig. 10.
Note again that we set $\epsilon=0.17$ when $\tilde t \le 1$ 
and $\epsilon=0.2$ when $\tilde t > 1$ for 
$M_0=150$ and 200 in order to avoid the initial convergence failure of
the Riemann solver.
So the injection is slower in these two models, resulting in a 
retarded evolution of all related quantities until $\tilde t > 1$.
In the evolved shocks the CR pressure at the subshock is higher for larger values of $M_0$ and
becomes $P_{\rm c,2} \approx 0.9 \rho_0 u_0^ {\prime2}$ for $M_0\gsim 50$.
For high initial Mach-number shocks ($M_0 \gsim 50$) the CR energy ratio 
approaches $\Phi \approx 0.6 $, resulting in $\Phi^\prime \approx \Phi
\cdot (|u_0|/u_0^\prime)^3 \approx 1$, 
since the approximate time-asymptotic value of $u_0^\prime/|u_0|$
is 0.83 for strong shocks.
This high efficiency is possible,
because CRs have absorbed some of the thermal and kinetic energy from
the initial shock structure inside the simulation box, in addition to
the kinetic energy that has entered the shock during the
simulation, $\int 0.5\rho_0 u_0^{\prime3} {\rm dt} $. 
We note that the ``undulating'' features in the time evolution of $P_{\rm c,2}$ seem to be
numerical artifacts, and not real.
Unlike other spatially averaged or integrated quantities, the plotted $P_{\rm c,2}$ 
is sampled exactly at the subshock (\ie from one zone) whose properties 
show small, noisy variations in time.
These features seem in particular to correspond to times when the 
subshock crosses a regular zone boundary and are most prominent for the
injection models with $\epsilon=0.17-0.2$. 

Once again the initial injection rate before the weakening of the
subshock is independent of $M_0$ for strong shocks, and it is 
about 4 times smaller than that of $\epsilon=0.25$ models and
about 8 times smaller than that of $\epsilon=0.3$ models.
For $1<\tilde t<200$, the time averaged injection rate can be fitted as a power-law,
$\xi\sim 2.5\times 10^{-3} {\tilde t}^{-0.4}$, 
for strong Mach shocks.
As discussed in \S5.1.2, the injection rate decreases with time
mainly because the subshock
weakens due to CR modification, but partly because 
the diffusive flux of injected particles crossing the shock 
decreases for smaller injection momentum.
If the effective numerical subshock thickness (spatial resolution) were set up to decrease 
as the postshock temperature decreased, we expect the injection rate for high
$M_0$ shocks would level off to the level similar to that for low $M_0$ shocks,
that is, $\xi \sim 10^{-3.4}$ found for the $M_0=5$ case. 
Below we will list this for our estimated asymptotic rate.
Considering the lack of our understanding of the detail physics of the
shock transition, this should be considered a rough estimate.

The same set of quantities are shown in the right three panels of Fig. 10 for $M_0=30$ 
for $\epsilon=0.2$, 0.23, and 0.25, in order to study the effects 
of different injection rates.
Obviously larger values of $\epsilon$ (\ie more leakage) lead to higher injection
rate, $\xi$, and so acceleration of CRs proceeds faster.
For all values of $\epsilon$ considered, however,
time-asymptotic values of $P_{\rm c,2}/\rho_0 (u_0\prime)^2$ 
$\sim 0.8$ and $\Phi\sim 0.6$, even though the injection histories are different.
This implies that the long term behavior of $\Phi$ is mostly
dependent on the initial shock Mach number, but only weakly dependent on 
the details of the injection rate, so long as it is
in the range seen here. If the injection rate falls
below some critical range, other studies have
emphasized that shock modification can be
minimal, so that ``test particle'' solutions
apply \citep[\eg][]{berz95, mdv00}. 
It seems mostly likely that all the shocks simulated here will remain
strongly modified, so correspond to ``high injection efficiency'',
although we cannot yet judge if they would eventually approach some
kind of self-organized critical condition such as that suggested
by \citet{mdv00}.
Evaluating that possible eventuality will likely require 
self-consistent treatments of the coupling between evolution of
CRs and the diffusion coefficient. That is beyond the scope of the
present work.

\subsubsection{The Particle Distribution} 
The evolution of the CR distribution function for a $M_0=100$ shock
is shown in Fig.~11.
The general behavior is the same as the early evolution 
for the $M_0=30$ shock with $\epsilon=0.3$ shown in Fig. 8. 
In particular both the thermal distribution and the transparency function shift to lower
momenta, the function $g(p)$ shows the
characteristic ``concave upwards'' curves reflecting modified shock structure.
The CR spectrum extends to higher momenta with time, 
but the amplitude at intermediate momenta decreases with time. 
The slope of the CR spectrum, defined as 
$q= - (\partial {\rm ln}f/ \partial {\rm ln} p)$,
ranges over $4.1 \lsim q \lsim 4.4$ near $p_{\rm inj}$, reflecting
$r_{sub} \gsim 3$ then decreases with the 
particle momentum and converges for strong shocks to $q \sim 3.1$ just below $p_{\rm max}$.

\subsection{Time Asymptotic Behaviors} 
Although our CRASH code is much more efficient than conventional codes
used for this problem,
we have so far integrated our models only until the maximum momentum 
reaches about $\sim 40m_{\rm p}c$ ($\tilde t =200$),
since computational requirements of using a Bohm type
diffusion model are still substantial.
However, some characteristics of the shock seem to have reached
quasi-steady states well before the end times of our simulations.
So we can make an attempt to project the long-term, 
time-asymptotic behaviors of these CR modified shocks from our results.
For strong shocks of $M_0 \gsim 30$ the CR injection via thermal leakage is
very efficient, and so the CR pressure quickly dominates over the gas
pressure.
The postshock CR pressure reaches approximate time-asymptotic values 
by the time the injection and acceleration become balanced by
the advection and diffusion of CRs away from the shocks. 
As compression increases through the precursor, the subshock
slows down, and the subshock velocity relative to the far upstream flow
approaches $u_0' \sim (0.83-0.85) |u_0|$ for strong shocks with the inverse wave
amplitude parameter considered here, $\epsilon= 0.2-0.3$.
As a result, the postshock CR pressure asymptotes to a quasi-steady
value of $P_{\rm c,2} \sim 0.9 \rho_0 u_0'^2\sim 0.63 \rho_0 u_0^2$ for strong shocks.
Approximate estimates for time-asymptotic values of the CR energy ratio, $\Phi$,
are calculated at $\tilde t=20$ for models with $\epsilon=0.3$ and 0.25, and 
at $\tilde t=100$ for models with $\epsilon=0.2$. 
They are plotted as a function of the initial shock Mach number $M_0$
in the right top panel of Fig. 12.
Also time-asymptotic values of $u_0^\prime/|u_0| = M_0^\prime/M_0$, 
are calculated at $\tilde t=100$ for models with $\epsilon=0.2$ and
shown in the right middle panel of Fig. 12.
We can estimate time asymptotic values of $\Phi^\prime$ as
$\Phi \cdot (|u_0|/u_0^\prime)^3$ using data in these two plots.
We also show time-asymptotic values of
the CR pressure at the shock normalized to the ram pressure of the
initial shock, $P_{\rm c,2}/\rho_0u_0^2$ (solid line), and that
normalized by the ram pressure of the instantaneous shock,
$P_{\rm c,2}/\rho_0u_0^{\prime2}$ (dotted line), which are calculated
at $\tilde t =100$ for $\epsilon=0.2$ shocks
in the right bottom panel of Fig. 12. 

Some other characteristics of the subshock, on the other hand,
show slow secular changes,
even after the postshock $P_c$ becomes quasi-steady. 
As the CR distribution continues to harden, for example,
the effective adiabatic index of the flow decreases with time.
Also the precursor spreads further out, the compression through the
precursor increases and the subshock continues to weaken. 
The entire shock structure broadens linearly with time in a ``self-similar''
way, because of the strongly momentum dependent diffusion matching the
advective spreading behind the shock.
As a result, unless we enforce an escaping upper momentum boundary
(\ie $g(p)=0$ for $p>p_1$, where $p_1$ is the upper momentum
boundary), the CR modified shock would not reach a strict steady-state.

According to kinetic equation simulations of spherical blast waves
\citep{berz95,dvb95}, and both Monte Carlo and kinetic equation 
calculations of steady planar shocks with strongly momentum
dependent diffusion
\citep{mal97,berzell99}, a CR modified shock characteristically
retains a subshock 
with the compression ratio $2.5< r_{\rm sub}< 4$
and does not become a completely smooth transition.
In the present time dependent planar
shock simulations with Bohm-like diffusion
we also found that the subshock remains.
The values
for the subshock Mach numbers in our shocks can be 
described by the scaling $M_s \sim 2.9 M_0^{0.13}$
at the end of the simulations.
Since the subshock compression ratio is related to the gas subshock
Mach number as $r_{\rm sub}= 4/(1+ 3/M_s^2)$, this leads to
$ 2.5< r_{\rm sub}< 4$.
Steady, adiabatic compression through the precursor
can be expressed simply as
$\rho_1/\rho_0 = (M_0^\prime/M_s)^{(\gamma_{\rm g}+1)/2}$,
where $M_0^\prime= M_0 \cdot (u_0^\prime /u_0)$ is the time asymptotic value of the
shock Mach number with respect to the far upstream flow.
Then the total compression ratio $r_{\rm tot} = \rho_2/\rho_0$ for a
{\it steady} CR modified shock with a subshock can be expressed
in terms of the subshock Mach number and the subshock compression 
ratio as
\begin{equation}
 r_{\rm tot} = r_{\rm sub} (M_0^\prime/M_s)^{3/4}
\end{equation}
\citep{berzell99}.
Here $r_{\rm sub} M_s^{-3/4} = 4M_s^{5/4}/(M_s^2 + 3)$ from the
shock jump condition. 
So, $r_{\rm sub} M_s^{-3/4} = 0.96-1.36$ for the subshock Mach
numbers found in our simulations (\ie $2\le M_s \le 6$). 
Both \citet{berz95} and \citet{berzell99} found that 
$ r_{\rm sub} M_s^{-3/4} \sim 1$ for total Mach numbers ($M_0^\prime$) 
considered, so that the total compression ratio 
scales with the total Mach number as $r_{\rm tot} \sim (M_0^\prime)^{3/4}$. 
For comparison we plotted 
in the left three panels of Fig. 12
the preshock ($\rho_1/\rho_0$), postshock ($\rho_2/\rho_0$) 
densities, and subshock compression factor ($r_{\rm sub}$)
at $\tilde t=20$ for models with $\epsilon=0.3$ and 0.25, and 
at $\tilde t=100$ for models with $\epsilon=0.2$ 
as a function of the initial shock Mach number $M_0$.
We note once again that the precursor compression continues, although very slowly,
so these quantities have not reached time-asymptotic values 
at the terminal time of our simulations. 
For our simulated shocks with $\epsilon=0.2$ the ratio $\rho_2/\rho_0$ shown in Fig. 12  
can be approximated by a power-law, $\rho_2 / \rho_0 \sim 1.5 M_0^{0.6}$ for $M_0 < 80$.
But this power-law relation flattens out for $M_0>80$.
Considering that $\rho_2$ has not reached a quasi-steady value at $\tilde t=20$ or
$\tilde t=100$, especially for high Mach number shocks, 
and that the ratio $M_o^\prime /M_0$ decreases with $M_0$,
our empirical relation for $r_{\rm tot}$ vs. $M_0^\prime$ is likely to 
steepen if we run the simulations for much longer times.
As discussed in \S 5.1.1, the scaling relation for the compression ratio
results simply from the fact that the subshock persists in our simulated shocks, 
which broadens in space self-similarly with time due to strongly
momentum dependent diffusivity, but are not yet stationary.
According to \citet{mal97}, the total compression saturates
at the level of $(\nu p_{\rm max}/p_{\rm inj})$ for $M_0^\prime > (\nu p_{\rm max}/p_{\rm inj})^{4/3}$, 
where $\nu$ is his injection parameter.
For strong shocks in our simulations, $\nu \sim 4 \times 10^{-2}$ and
$p_{\rm max}/ p_{\rm inj} \sim 2\times 10^4$, 
so the shocks considered here are not in
the saturation limit.

\section{Summary}

In order to study cosmic-ray modified shocks we have developed a new numerical code, 
CRASH (Cosmic-Ray Amr SHock), 
by implementing a thermal leakage injection scheme introduced
by \citet{gies00}
into a new hydro/CR code with Adaptive Mesh Refinement and Shock
Tracking scheme by \citet{kang01}.
Our CR injection model is based on the ``thermal leakage'' process 
at quasi-parallel CR shocks. Injection is regulated by the 
convolution of the population in the high energy tail of the 
Maxwellian velocity distribution of the postshock gas and 
a transparency function, $\tau_{\rm esc}(p,u_d,\epsilon)$,
determined by the strength of downstream MHD waves, expressed
by a parameter, $\epsilon$.
The injection rate in our model is then controlled largely by the subshock Mach number, $M_s$,
since that determines the ratio of the postshock flow speed to the
breadth of the thermal distribution. For weak shocks injection
becomes more difficult as $M_s$ decreases, but is independent of $M_s$ for 
strong shocks, when the subshock compression asymptotes.
With our CRASH code the CR injection and acceleration at astrophysical
shocks can be simulated numerically even for strongly momentum dependent
spatial diffusion coefficients. 

Using these tools the time evolution of CR shocks has been 
followed with a Bohm type diffusion model. 
We started from the initial conditions for pure gasdynamic shocks of various 
initial Mach numbers ($M_0$) without any pre-existing CRs.
In such simulations the CR injection rate is not treated as a
fixed free parameter, as it has been done traditionally.
Instead, once an ``intelligent estimate'' is made of the
strength of the trapping wave field of the downstream MHD
turbulence, the injection is followed naturally and
self-consistently during nonlinear evolution of the flows.
For strong shocks with initial strengths, $M_0\gsim30$, 
a substantial portion of the particles in the tail of the Maxwellian 
distribution have
velocities high enough to leak upstream against the wave-particle
interactions, so the injection is efficient and fast.
As the CR pressure increases at the subshock and the 
in-flowing plasma is compressed, however, the subshock slows down with respect
to the far-upstream flow. This modification occurs rather promptly 
and before 
the postshock CR pressure reaches an approximate time-asymptotic value.  
Afterwards the evolution of the shock structure becomes secular.
For strong shocks, the subshock speed in our
simulations decreases by about 15-17 \%,
and the postshock CR pressure absorbs up to 60 \% of the ram pressure of 
the initial shock, which corresponds to 90 \% of the ram pressure of the 
upstream flow in the evolved subshock rest frame at the end of our
simulations.
Once the postshock CR pressure becomes constant, 
the shock structure evolves approximately in a ``self-similar'' way,
because the scale length of shock broadening increases linearly with time. 

The injection rate, defined as the fraction of the particles
passed through the shock that are accelerated to form the CR population,
becomes as high as $\xi \sim 0.01$ early in the evolution of strong shocks, 
independent of $M_0$.
As the shock is modified, however,
the subshock Mach number decreases down to $M_s\sim 2.9 M_0^{0.13}$,
and the injection rate reduces towards the limiting value corresponding 
to weak subshocks.
Our approximate numerical estimate for the limiting value is 
$\xi \sim 10^{-3.4}$ for models with $\epsilon=0.2$. 
Since the CR spectrum continues to extend to higher momenta,
and since those highest energy particles diffuse
rapidly away from the shock when $\kappa \propto p$,
energy begins to be transported away from the
shock transition. 
This allows the subshock to persist and
the total compression to become large compared to steady energy-conserving
gasdynamic shocks. 
Also the shock structure evolves towards the high
compression ratios seen in kinetic equation
spherical CR shocks and steady state planar
shocks with energy escape through spatial or momentum boundaries.

Finally, the main conclusions of our time-dependent simulations of CR modified shocks
based on Bohm-like diffusion and
a physically-based thermal leakage CR injection are:

1. In the strong shock limit of initial Mach number $M_0 \gsim 30$,
significant physical processes such as injection and acceleration
become largely independent of the initial shock Mach number,
and only weakly depend on the postshock wave amplitude parameter, $\epsilon$, for values considered
here ($0.2\le \epsilon \le0.3$).
According to our thermal leakage model, the overall injection rate 
approaches $\xi \sim 10^{-3}$ and the fraction of upstream flow kinetic energy 
{\it in the initial shock rest frame} that has been transferred 
to CRs is $\Phi \sim 0.6$  for strong shocks. 
The ratio of the postshock CR pressure to the {\it instantaneous}
ram pressure of the subshock with respect to the upstream plasma
approaches $\sim 90$\%.
On the other hand, the injection rate and acceleration efficiency are sensitively 
dependent on $M_0$ for low Mach number shocks ($M_0\lsim 30$).

2. In our simulations, with no initial CRs around the shock,
the thermal leakage injection is very efficient initially 
when the subshock is strong, but it becomes much less efficient as the subshock weakens
due to nonlinear feedback from the CR pressure. 
For example, the time-averaged injection rate 
can be fitted as a power-law form, $\xi \sim 2.5\times 10^{-3} ( \tilde t)^{-0.4}$
for $1<\tilde t< 200$ ($\tilde t = t/t_0$) for strong shocks of $M_0\gsim 30$
and the inverse wave-amplitude parameter $\epsilon =0.2$.

3. Although some particles injected early in the
shock evolution continue to be accelerated to ever higher energies, 
the immediate postshock CR pressure reaches a quasi-steady value
when a balance between injection/acceleration and advection/diffusion
is achieved. The region of quasi-steady postshock properties
spreads for Bohm diffusion in an almost self-similar fashion
because both diffusive and advection rates scale linearly
with time. This expansion maintains the large precursor compression
at values that can be large compared to what one would derive
from Rankine-Hugoniot relations for steady gas shocks with a
relativistic equation of state.

4. The sum of the compression through the precursor and across the subshock
calculated near the terminal time, $\tilde t=100$, for models with $\epsilon=0.2$
can be approximated by $\rho_2 / \rho_0 \sim 1.5 M_0^{0.6}$ for $M_0 < 80$,
but this $\rho_2/\rho_0 - M_0$ relation flattens for $M_0\ge 80$. 
Considering that $\rho_2$ continues to increase until the terminal time,
especially for higher Mach number shocks, and that the Mach number of the 
subshock with respect to the far upstream ($M_0^\prime$) decreases 
for strong shocks with high initial Mach numbers ($M_0$), 
this is probably consistent with  the
$\rho_2/\rho_0 \propto M_0^{\prime 3/4}$ scaling relation found
for steady shocks and for spherical shocks
by several authors \citep{berz95, mal97, berzell99}.
The large density compression is possible in the planar shocks without energy
escape from the whole computed system, 
since the shock structure  spreads out far upstream and far
downstream, due to strong momentum dependent diffusivity. 

Although our simulations have been carried out only until the highest momentum
is $p_{\rm max} \sim 40 m_{\rm p}c$ due to severe computational requirements, these simulation
results provide useful guidance for long-term evolution of CR modified shocks.
In a future study, we will extend the integration time so that 
$p_{\rm max}/m_{\rm p}c\gg1$
can be achieved and the free escape at the upper momentum boundary is applied
in order to compare the time-dependent simulations with previous studies of
steady-state shocks \citep{elleich85,mal97,madru01}.

\acknowledgements
The authors would like to thank D. Ryu and M. Malkov for helpful discussions.  
HK was supported by Korea Research Foundation Grant (KRF-2000-015-DP0448).
TWJ is supported by NSF grant AST-9619438, by NASA grant NAG5-10774
and by the University of Minnesota Supercomputing Institute, 
The numerical calculations were performed 
through "The 2nd Supercomputing Application
Support Program" of the KISTI (Korea Institute of Science and Technology 
Information).

\appendix
\section{Discussion of Test Calculations}

The algorithms applied in CRASH are already mostly well-documented and 
well-tested. The hydrodynamic scheme, for example, is a member of
the well-known Godunov family and was verified by \citet{levshy95}.
Our Crank-Nicholson scheme for solution of the diffusion-convection 
equation was tested against previous, independent 
implementations and analytic test particle diffusive shock
acceleration solutions in \citet{kanjon91}, where we employed the PPM 
hydrodynamics algorithm. Other tests of nonlinear CR shock solutions 
using several previous hydrodynamical implementations were given for both
diffusion-convection \citep[\eg][]{kjr92,kanjon95} and
two-fluid \citep[\eg][]{jk90,kanjon95,kanjon97} models of CR transport. 
\citet{kanjon95} using PPM hydrodynamics and \citet{kanjon97} using 
TVD magnetohydrodynamics methods, for example,
provided comparisons between the CR transport approach applied here 
with our earlier ``thermal leakage'' injection model
and Monte Carlo, hybrid plasma simulations, and {\it in situ}
measurements of heliospheric shocks. While the newer
injection scheme utilizes a significantly improved physical model
to determine which particles are injected at a shock, both the
new and the old underlying numerical schemes have much in common, as 
discussed in \citet{gies00}. More recently \cite{kang01}
provided convergence tests of the combined shock tracking, AMR, convection
diffusion code upon which CRASH is based, again using our previous thermal 
leakage injection model.

In order to extend the above test base
to CRASH we present here several comparisons of
solutions to results from these earlier works.
In particular we have tested CRASH against piston-driven shock problems
that were calculated with our PPM/CR code \citep{jk90, kjr92}.
First, using a two-fluid version of our CRASH code with the modified 
entropy equation, we calculated a shock driven by a constant speed, plane 
piston with $u_p=0.3$ moving into a medium of $\rho=1$, $P_g=7 \times 10^{-4}$,
and $P_c= 3.5\times 10^{-4}$. Closure parameters are assumed to be
$\gamma_c=5/3$ and $<\kappa>=0.01$.
There is no injection of new CRs in this case, only reacceleration of
the previous CR population. Thus, it is a test for nonlinear
modified shocks of the dynamical
coupling terms included in equations (\ref{mocon})-(\ref{econ}).
Fig. 13 shows the evolution of the shock at $t=3,~6,$ and 9, which can
be compared with Fig. 1 of \citet{jk90}. Agreement is excellent.
The dashed horizontal line shows the steady state solution for $P_c$, 
which is fully consistent with the CRASH simulation.
This test calculation demonstrates that the
two fluid version of the CRASH code can accurately follow the evolution
of a strongly CR modified shock. The dynamical coupling between
the CRs and the gas is the same in the two-fluid model as in
the diffusion-convection, kinetic equation model, except for
the multi-scale character introduced by a momentum dependent
diffusion coefficient into the kinetic equation model.

To extend the dynamical evaluation to include momentum
dependent influences, we calculated another piston driven shock 
using the diffusion-convection version of CRASH. In this
case the piston has velocity $u_p=1$ moving
into a medium of $\rho=1$, $P_g=P_c=0.001$.
Two simulations were done, one with and one without
the modified entropy equation (\ref{scon}). 
The injection process was turned off
and only the base-grid was used (\ie $l_{\rm g,max}=0$). The diffusion
coefficient was given the form $\kappa=p^{0.5}$.
Fig. 14 shows the evolution of the shock at $t=10,~30,~50,~100$ and 150,
which was calculated with the modified entropy equation.
The last of these compares directly with the dotted curves in
Fig. 3 of \citet{kjr92}.
Some details of the early evolution differ, which cause the exact
locations and peaks of the transient density spike to differ
somewhat, but the overall structures and the shock transitions
at $t = 150$ are in very good agreement, including the compression,
as well as the gas and CR pressures.
The noted small early evolution differences among different codes have
been seen before. They reflect the sensitivity to numerical
details of the very rapid changes taking place in
the shocks at the start of such simulations.

Finally, we demonstrate the benefits of including the modified
entropy equation in treatments of strongly modified CR shocks. 
Fig. 15 shows from two simulations the subshocks and precursor structure 
for the piston-driven shock shown in Fig. 14 at $t=20$. The
two sets of curves were calculated
with the modified entropy condition enforced (S code: solid line) 
and without it (E code: dashed line).
The lower left panel shows the specific entropy, $\log{P_g/\rho^{\gamma}}$.
To the right of the subshock, where the flow is adiabatic, this
quantity should be a constant. However, because the flow there is
highly supersonic, the total energy is dominated by the kinetic energy.
Consequently, errors in computing the gas pressures used in
equations (\ref{mocon}) and (\ref{econ}), as 
extracted from a combination of the total energy and momentum
quantities lead to significant errors in the entropy of the
preshock gas. That in turn, leads to errors in the gas subshock and
CR acceleration behaviors. Enforcing entropy conservation in
smooth flows eliminates the problem, as illustrated in the figure.

\clearpage

\begin{figure}
\plotone{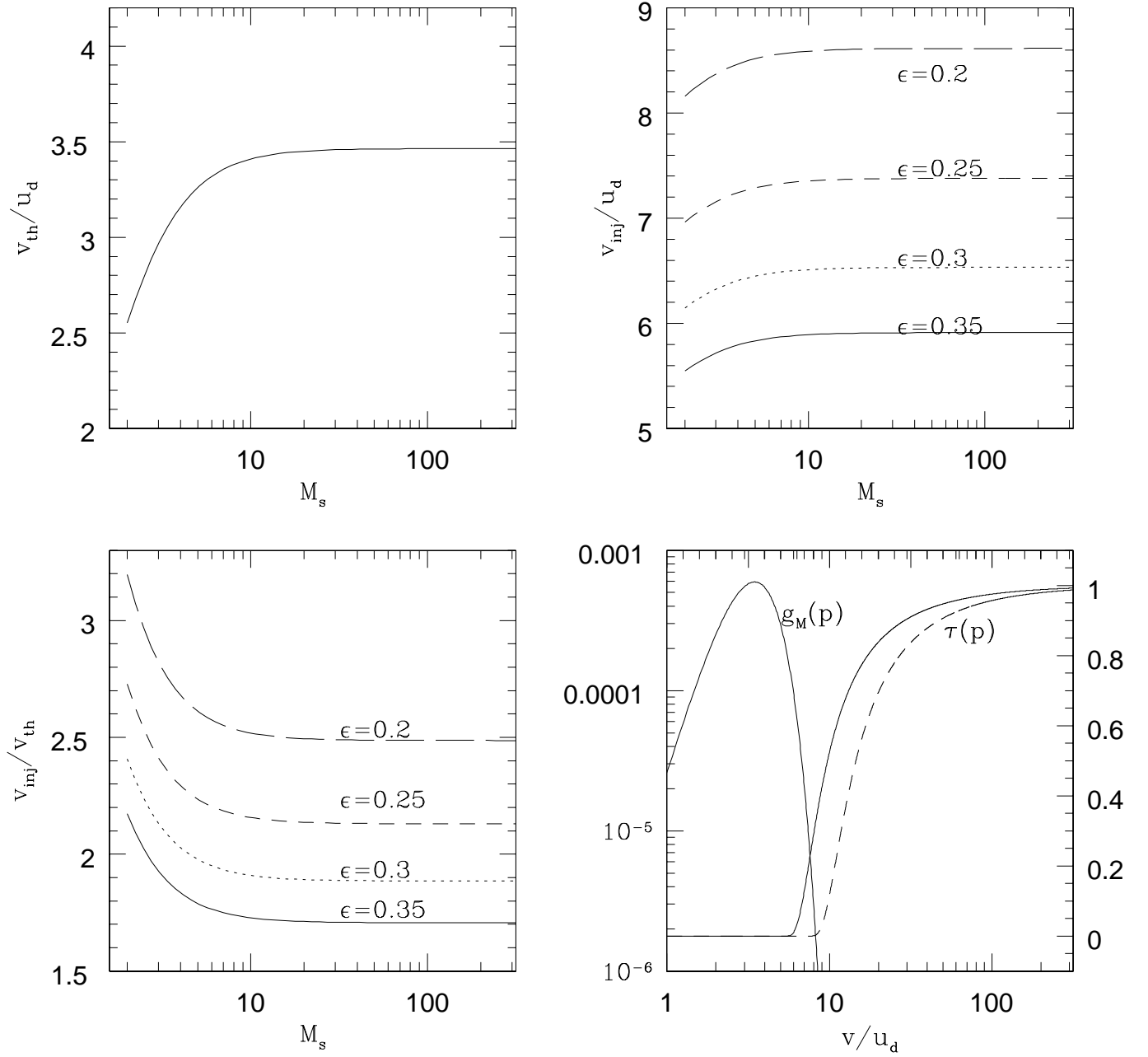}
\figcaption{ 
Thermal velocity, $\upsilon_{\rm th}$, and ``effective'' injection velocity, 
$\upsilon_{\rm inj}$, both normalized by the downstream flow velocity 
relative to the subshock, $u_d$, are shown as functions of subshock 
Mach numbers.
The effective injection velocity is also dependent on the inverse wave 
parameter, $\epsilon$.
In the lower left panel, the ratios of $\upsilon_{\rm inj}/\upsilon_{\rm th}$
are plotted for $\epsilon=0.2-0.35$.
Thermal leakage is faster for smaller $\upsilon_{\rm inj}/\upsilon_{\rm th}$.
In the lower right panel, the thermal distribution function, $g_{\rm M}$
of the downstream gas for $M_0=100$ shock is plotted 
against $\upsilon/u_d$, where $u_d= u_s/r_{\rm sub}$, $u_s/c=0.01$,
$c$ is the speed of light, and $r_{\rm sub}$ is the subshock compression ratio. 
The transparency function, $\tau_{\rm esc}$, for the same shock 
is also shown for $\epsilon = 0.3$ (solid line) and 0.2 (dashed line).
The axis on the left side is for $\log(g_{\rm M})$, while the axis
on the right side is for $\tau_{\rm esc}$. 
\label{fig1}}
\end{figure}
\clearpage

\begin{figure}
\plotone{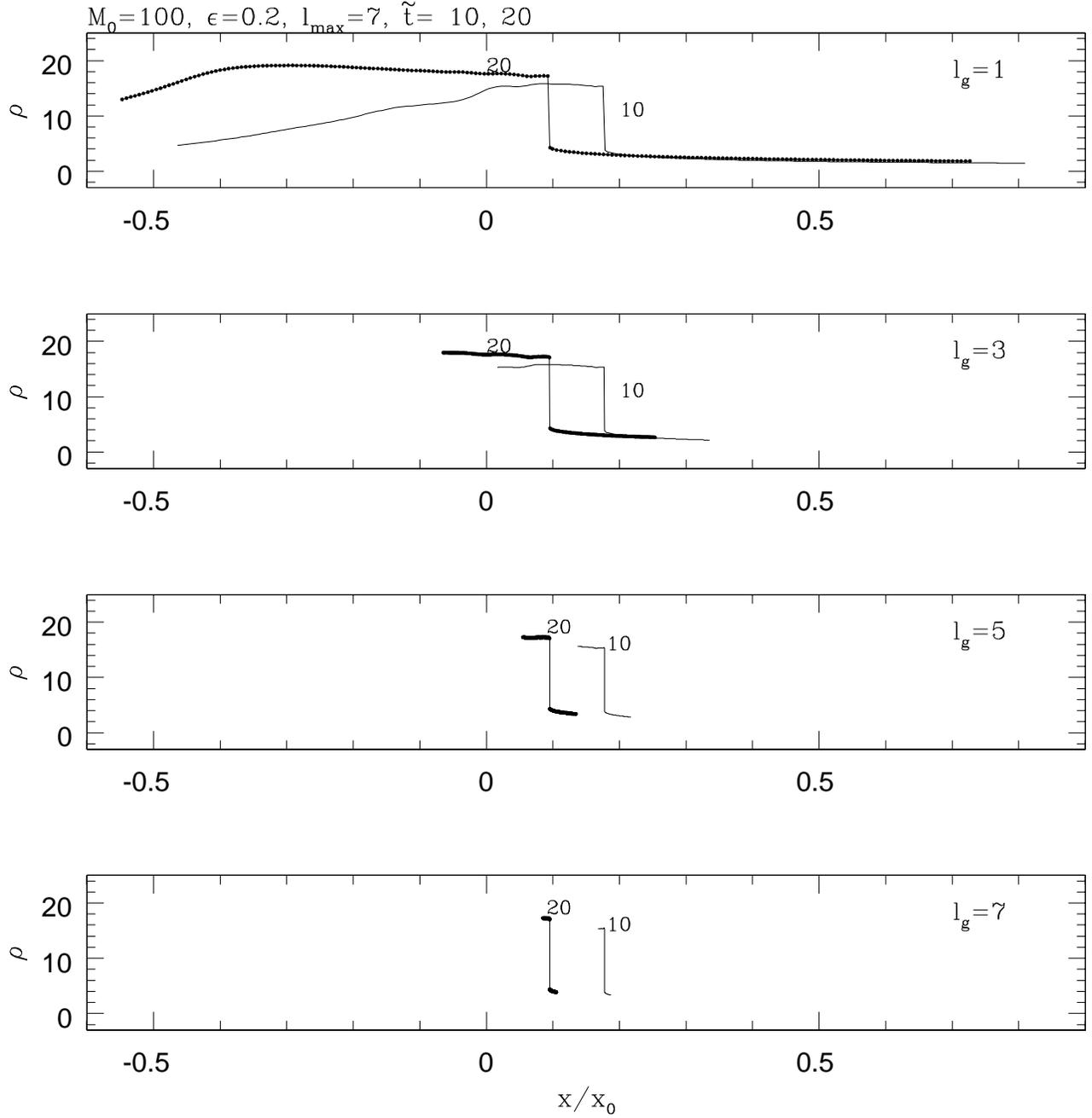}
\figcaption{
Density profile in the refinement region around the shock for $M_0=100$
and $\epsilon=0.2$ model at $\tilde t=10$ and 20 at the grid level $l_g=1,$ 3,
5, and 7. There are 200 zones at each grid level which are shown as filled 
points at $\tilde t=20$ curves.  
\label{fig2}}
\end{figure}
\clearpage

\begin{figure}
\plotone{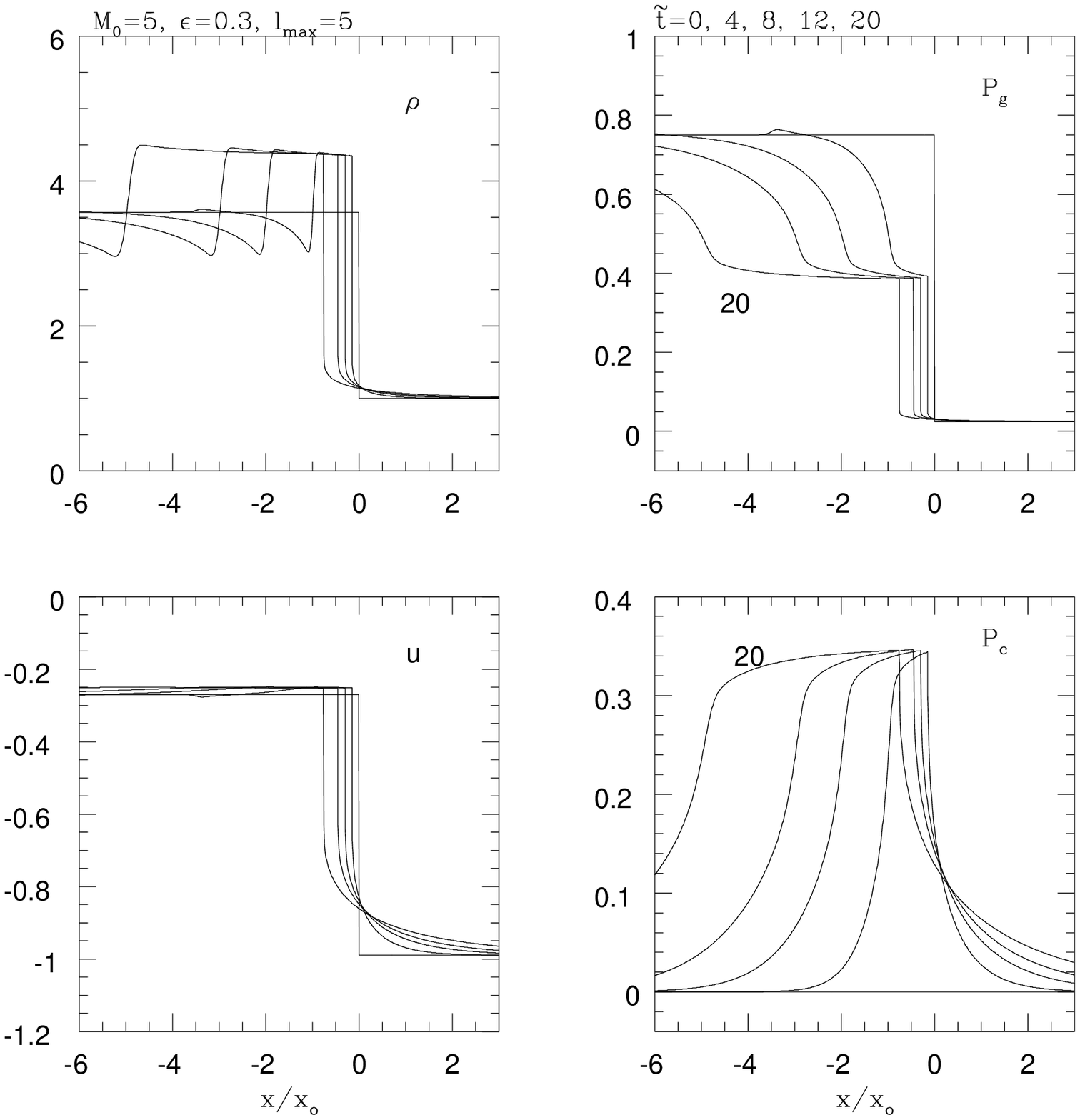}
\figcaption{
Time evolution of the $M_0=5$ shock structure with $\epsilon=0.3$ and
$l_{\rm max}=5$ refined grid levels.
Gas density, pressure, flow velocity and CR pressure are shown
at $t/t_{\rm o}=0,$ 4, 8, 12, and 20.
The right-facing shock drifts to the left, so the right most transitions
correspond to the earliest time $t=0$.
\label{fig3}}
\end{figure}
\clearpage

\begin{figure}
\plotone{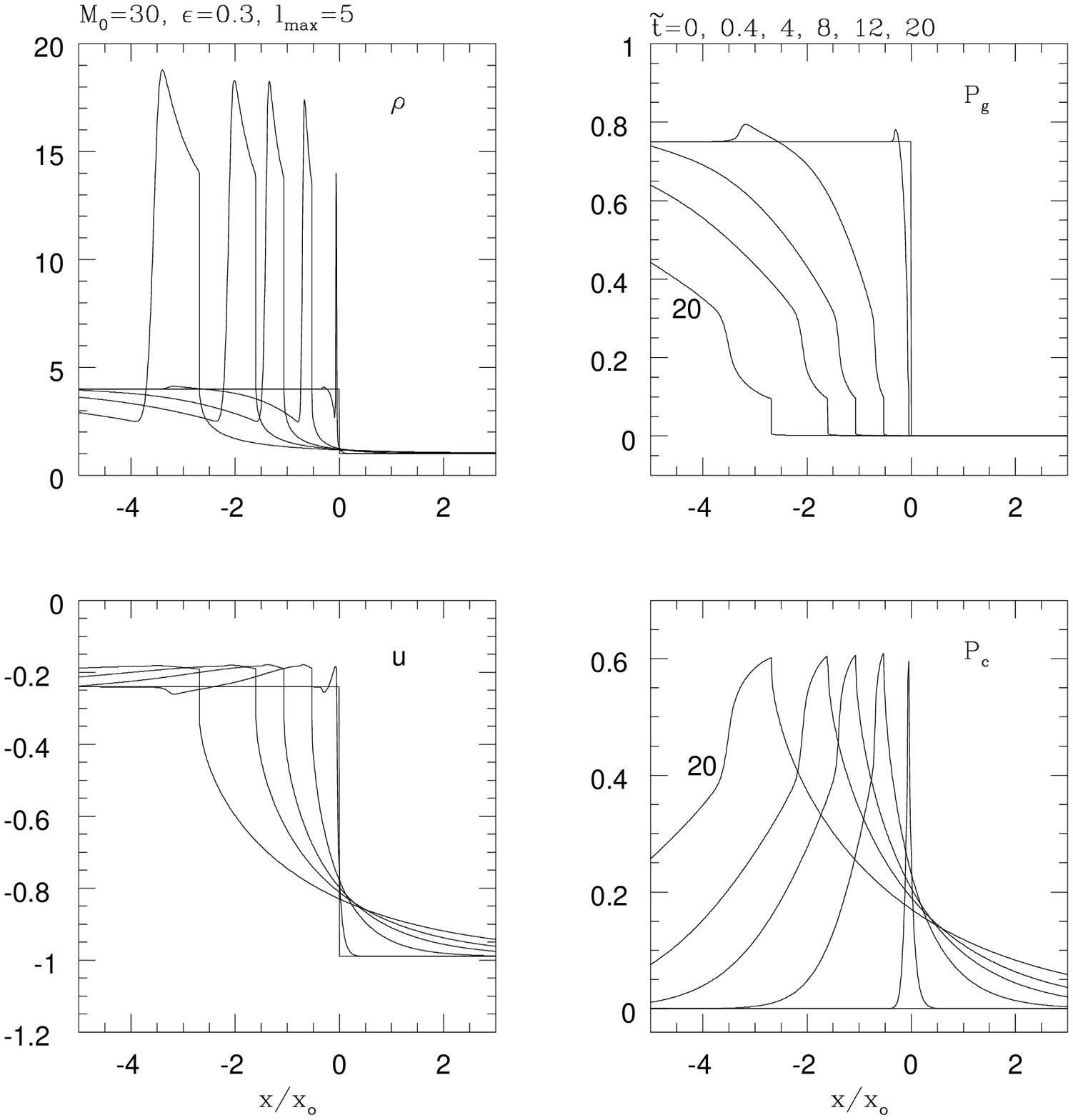}
\figcaption{
Time evolution of the $M_0=30$ shock structure with $\epsilon=0.3$ and
$l_{\rm max}=5$ refined grid levels.
Gas density, pressure, flow velocity and CR pressure are shown
at $t/t_{\rm o}=0,$ 4, 8, 12, and 20.
The shock drifts to the left, so the right most transitions correspond to the
earliest time $t=0$.
\label{fig4}}
\end{figure}
\clearpage

\begin{figure}
\plotone{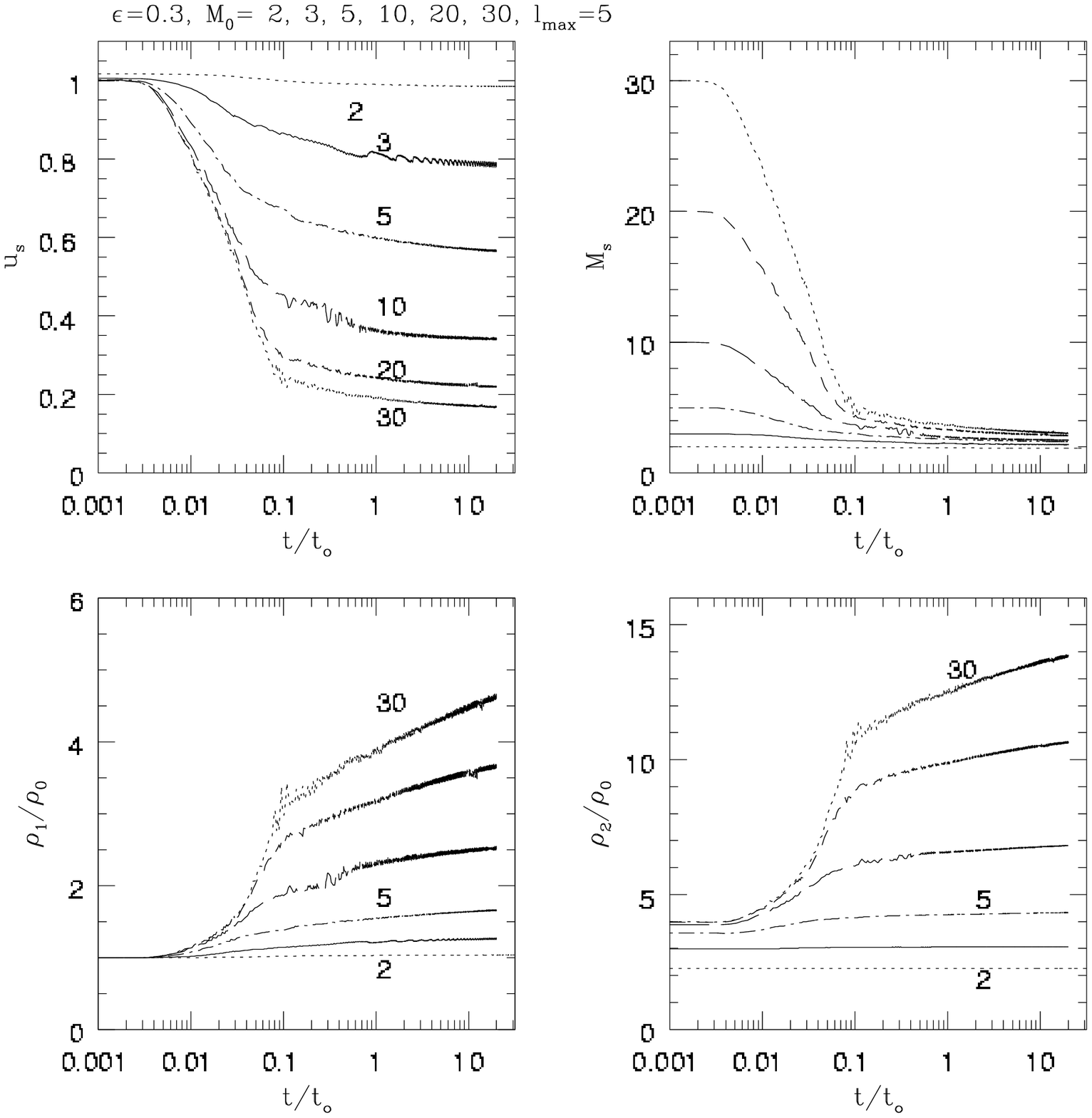}
\figcaption{
Time evolution of subshock velocity, $u_s$, subshock Mach number, $M_s$, 
preshock density,
$\rho_1/\rho_0$ and postshock density, $\rho_2/\rho_0$
are shown for $M_0=2-30$ and $\epsilon=0.3$.
\label{fig5}}
\end{figure}
\clearpage

\begin{figure}
\plotone{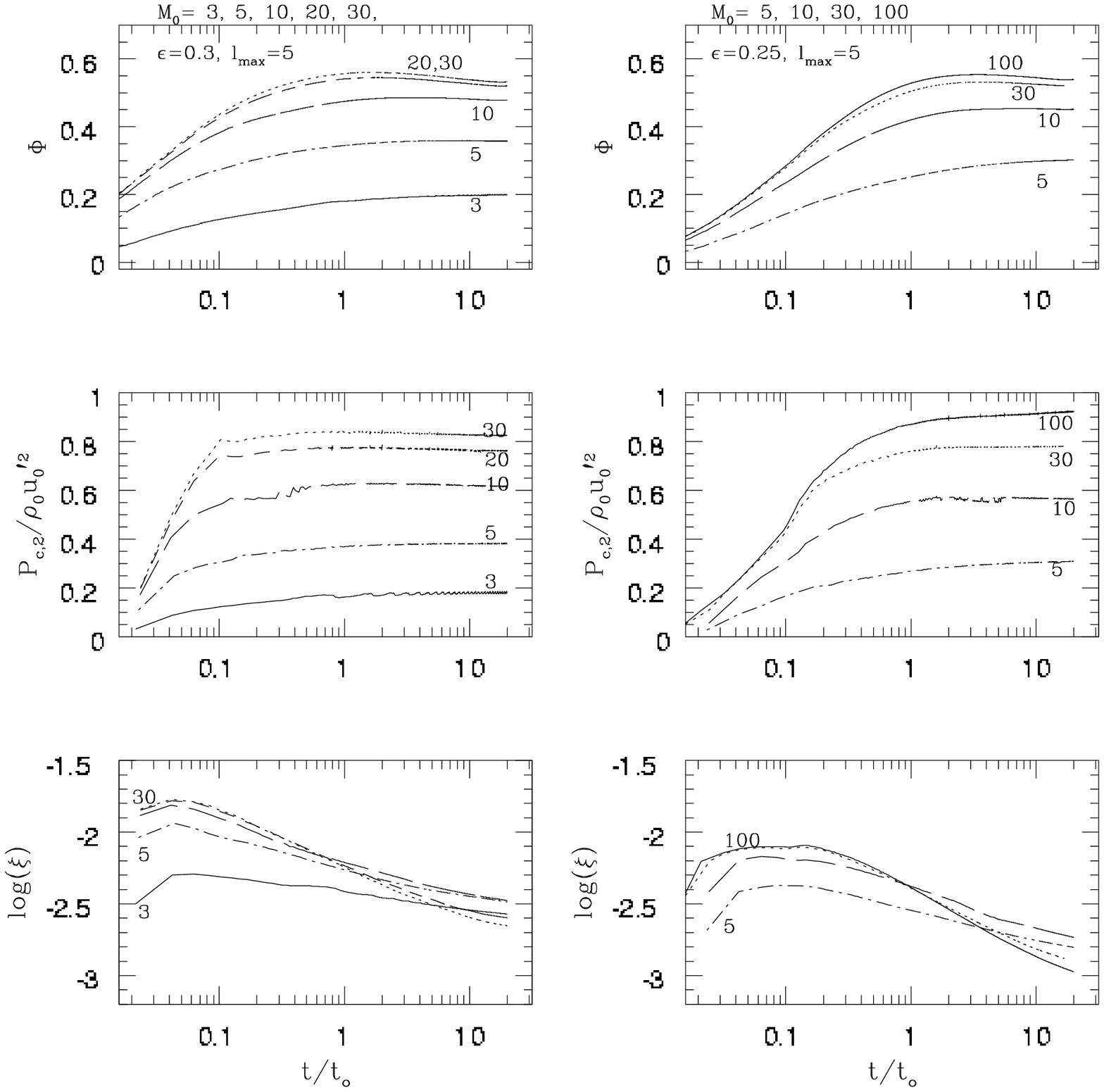}
\figcaption{
The ratio of total CR energy in the simulation box 
to the kinetic energy in the initial shock rest frame
that has entered the simulation box from upstream, $\Phi(t)$, 
the postshock CR pressure in units of upstream ram pressure in the instantaneous
shock frame, and time-averaged injection efficiency, $\xi(t)$.
Left three panels are for $M_0=3-30$ and  $\epsilon=0.3$.
Right three panels show the same quantities for $M_0=5-100$ and  $\epsilon=0.25$.
\label{fig6}}
\end{figure}
\clearpage

\begin{figure}
\plotone{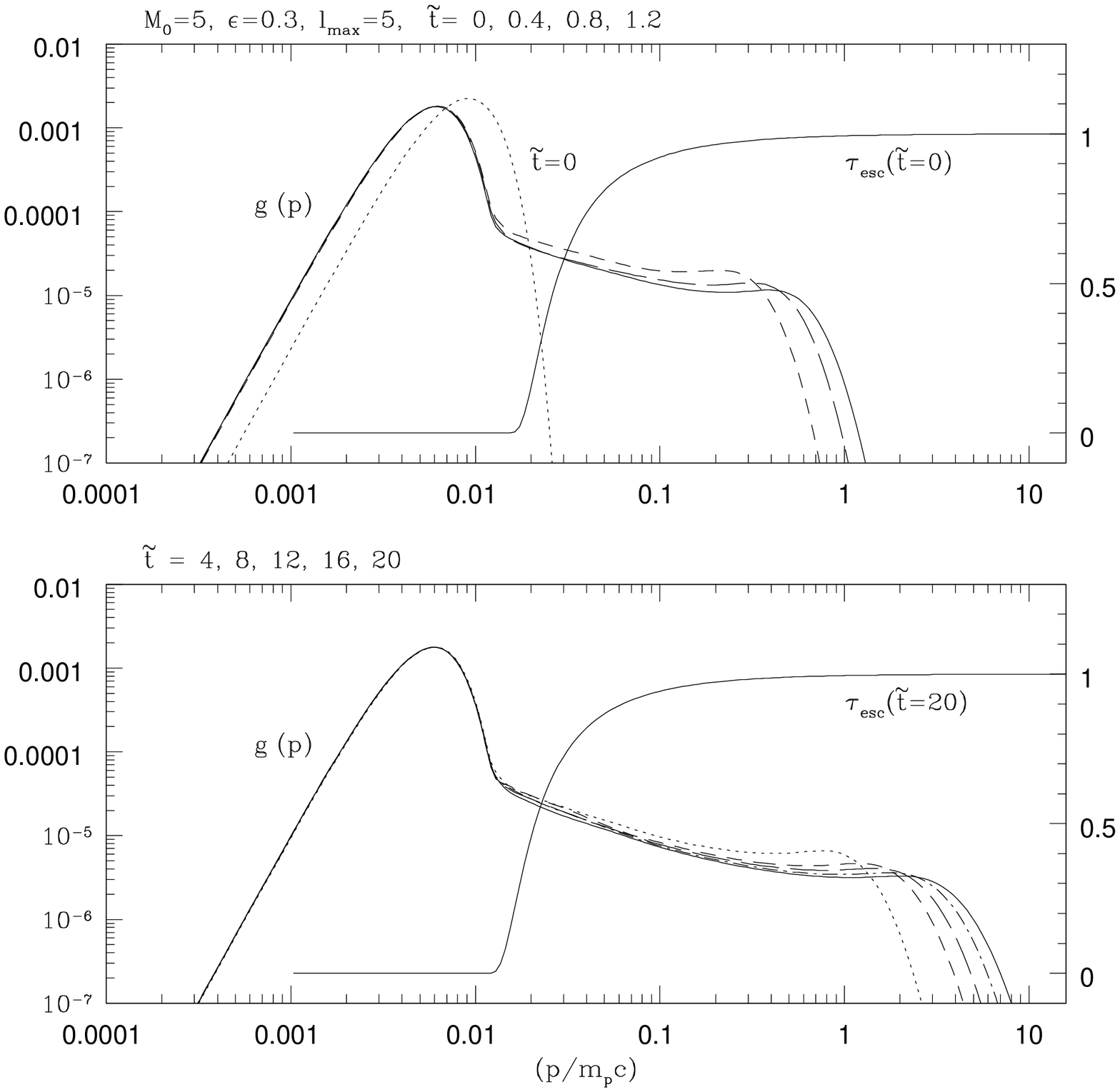}
\figcaption{
Evolution of the CR distribution function at the shock, represented
as $g=p^4f(p)$, is shown for the same shock as in Fig. 3 ($M_0=5$
and $\epsilon=0.3$). 
The transparency function is also plotted at $t/t_{\rm o}=0$ and 20 for reference.
The axis on the left side is for $\log(g_{\rm M})$, while the axis
on the right side is for $\tau_{\rm esc}$. 
\label{fig7}}
\end{figure}
\clearpage

\begin{figure}
\plotone{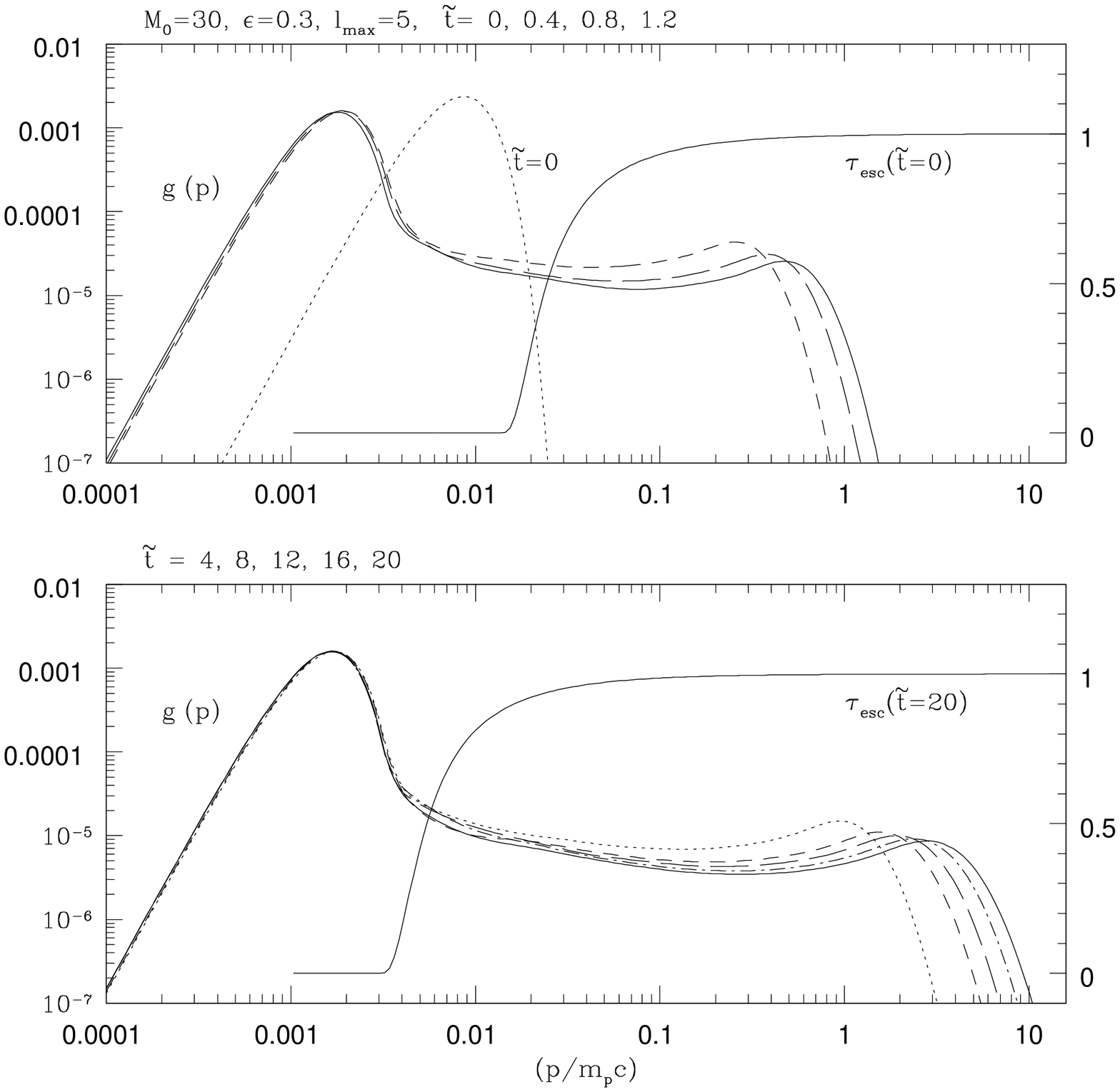}
\figcaption{
Evolution of the CR distribution function at the shock, represented
as $g=p^4f(p)$, is shown for the same shock in Fig. 4 ($M_0=30$
and $\epsilon=0.3$). 
The transparency function is also plotted at $t/t_{\rm o}=0$ and 20 for reference.
The axis on the left side is for $\log(g_{\rm M})$, while the axis
on the right side is for $\tau_{\rm esc}$. 
\label{fig8}}
\end{figure}
\clearpage

\begin{figure}
\plotone{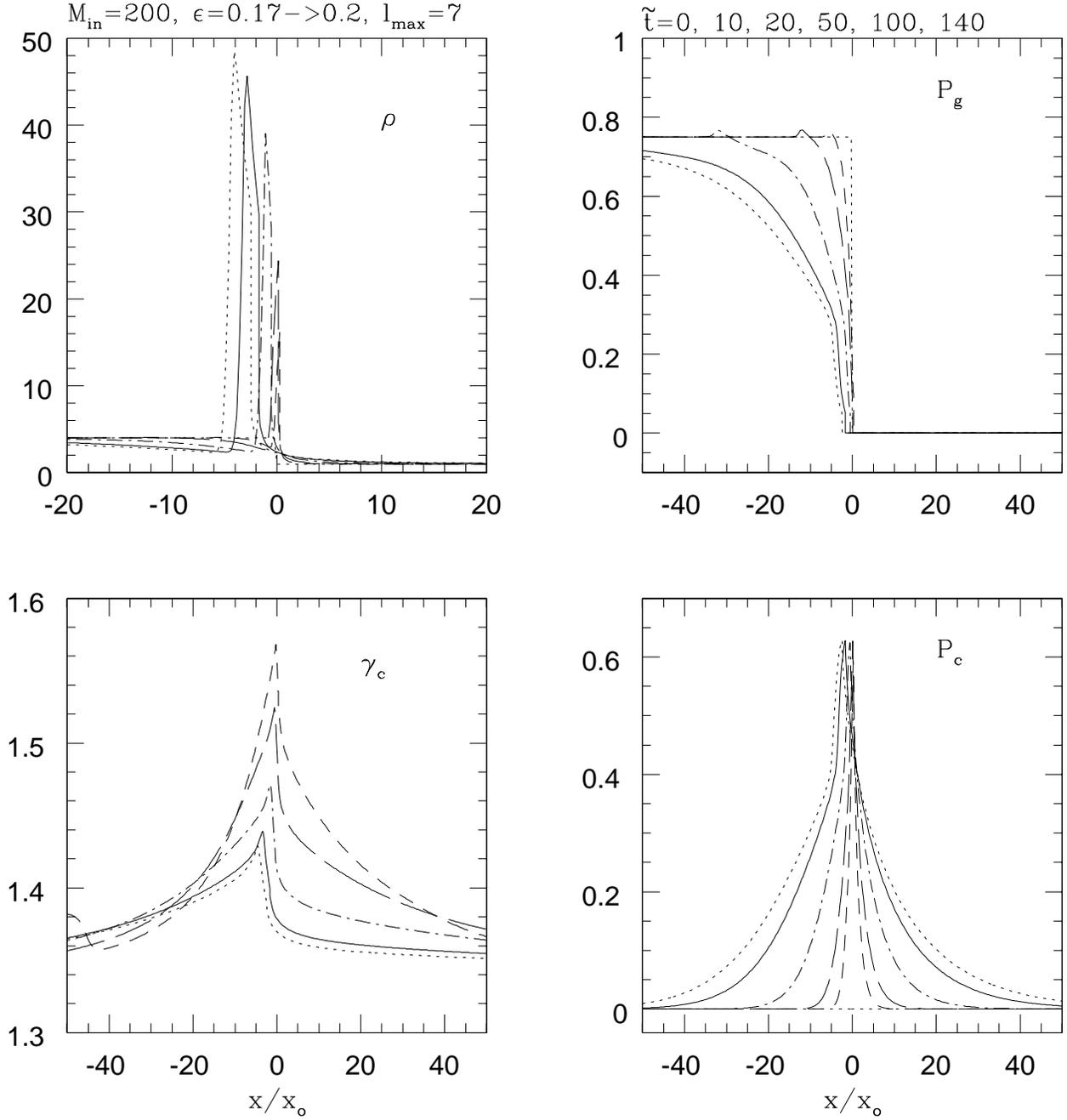}
\figcaption{
Gas density, pressure, CR adiabatic index and CR pressure are shown
at $t/t_{\rm o}=0,$ 10, 20, 50, 100, and 140 for a $M_0=200$ shock. 
The maximum level of refinement was $l_{\rm max}=7$.
The shock drifts to the left, so the right most transitions correspond to the
earliest time $t=0$.
\label{fig9}}
\end{figure}
\clearpage

\begin{figure}
\plotone{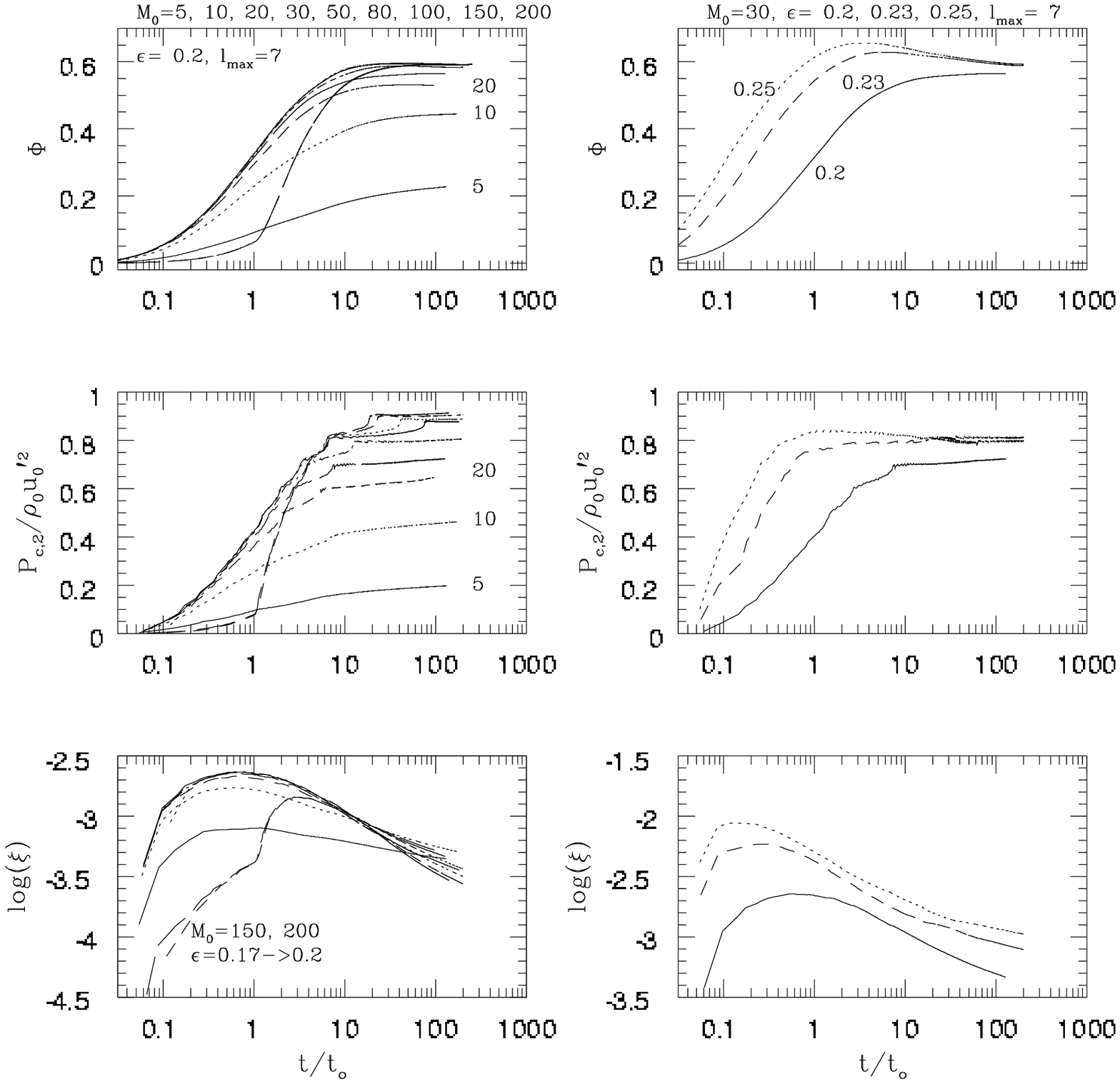}
\figcaption{
The ratio of total CR energy in the simulation box
to the kinetic energy in the initial shock rest frame
that has entered the simulation box from upstream, $\Phi(t)$, 
the postshock CR pressure in units of upstream ram pressure in the instantaneous
shock frame, and time-averaged injection efficiency, $\xi(t)$.
Left three panels are for $M_0=5-200$ and $\epsilon=0.2$.
Right three panels are for $M_0=30$ and $\epsilon=0.2$, 0.23, and 0.25. 
\label{fig10}}
\end{figure}
\clearpage

\begin{figure}
\plotone{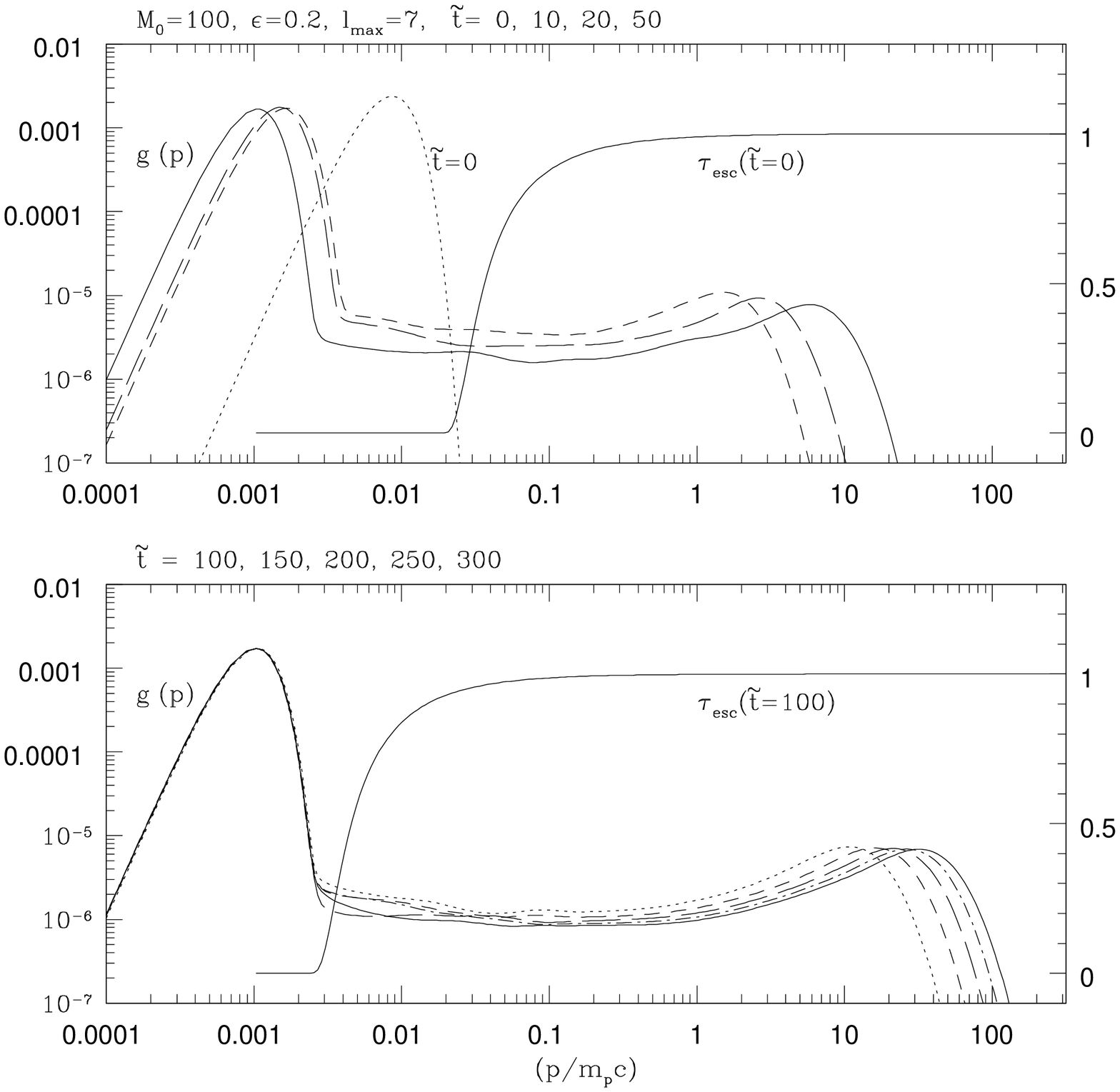}
\figcaption{
Evolution of the CR distribution function at the shock, represented
as $g=p^4f(p)$, is shown for the $M_0=100$ shock with $\epsilon=0.2$. 
The transparency function is also plotted at $t/t_{\rm o}=0$ and 100 for reference.
The axis on the left side is for $\log(g_{\rm M})$, while the axis
on the right side is for $\tau_{\rm esc}$. 
\label{fig11}}
\end{figure}
\clearpage

\begin{figure}
\plotone{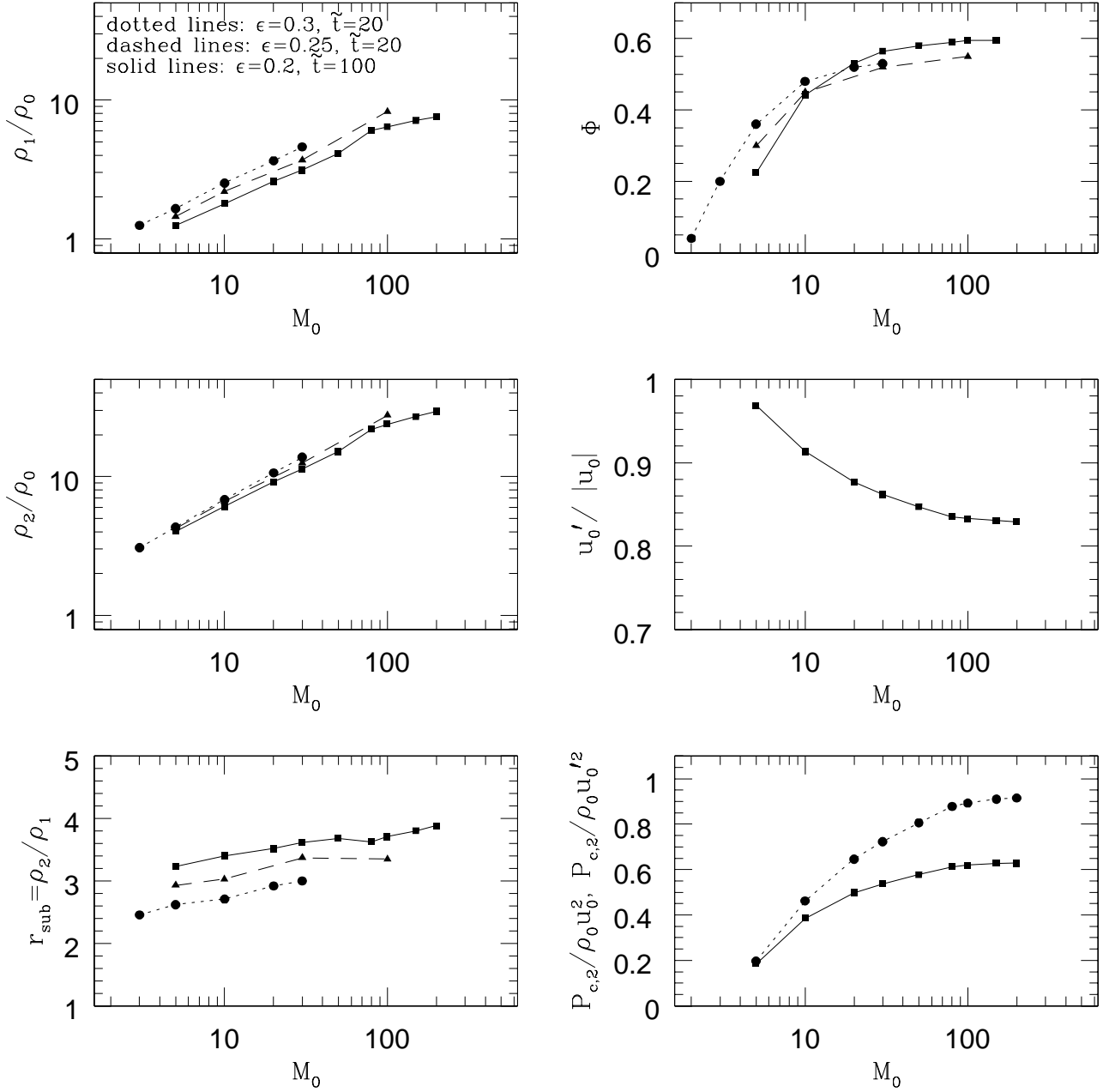}
\figcaption{
Left three panels:
preshock density, $\rho_1/\rho_0$,
postshock density, $\rho_2/\rho_0$,
and subshock compression ratio
for $M_0=5-200$.
For models with $\epsilon=0.3$ and 0.25 the numerical results
at $\tilde t=20$ are plotted, while for models with $\epsilon=0.2$
the results at $\tilde t=100$ are plotted.
Right top panel: approximate time asymptotic values of $\Phi$ is plotted.
The same symbols and line types are used as in $\rho_1/\rho_0$ plot.
Right middle panel: time asymptotic values of the upstream flow velocity relative
to the subshock, $u_0^\prime$, at $\tilde t=100$ are plotted for $\epsilon=0.2$.
Right bottom panel: time asymptotic postshock CR pressure normalized to
the ram pressure of the upstream flow in the initial shock rest frame, 
$P_{\rm c,2}/\rho_0u_0^2$ (solid line), 
and to the ram pressure of the upstream flow in the instantaneous shock rest frame, 
$P_{\rm c,2}/\rho_0u_0^{\prime2}$ (dotted line), 
is plotted for $\epsilon=0.2$. 
\label{fig12}}
\end{figure}
\clearpage

\begin{figure}
\plotone{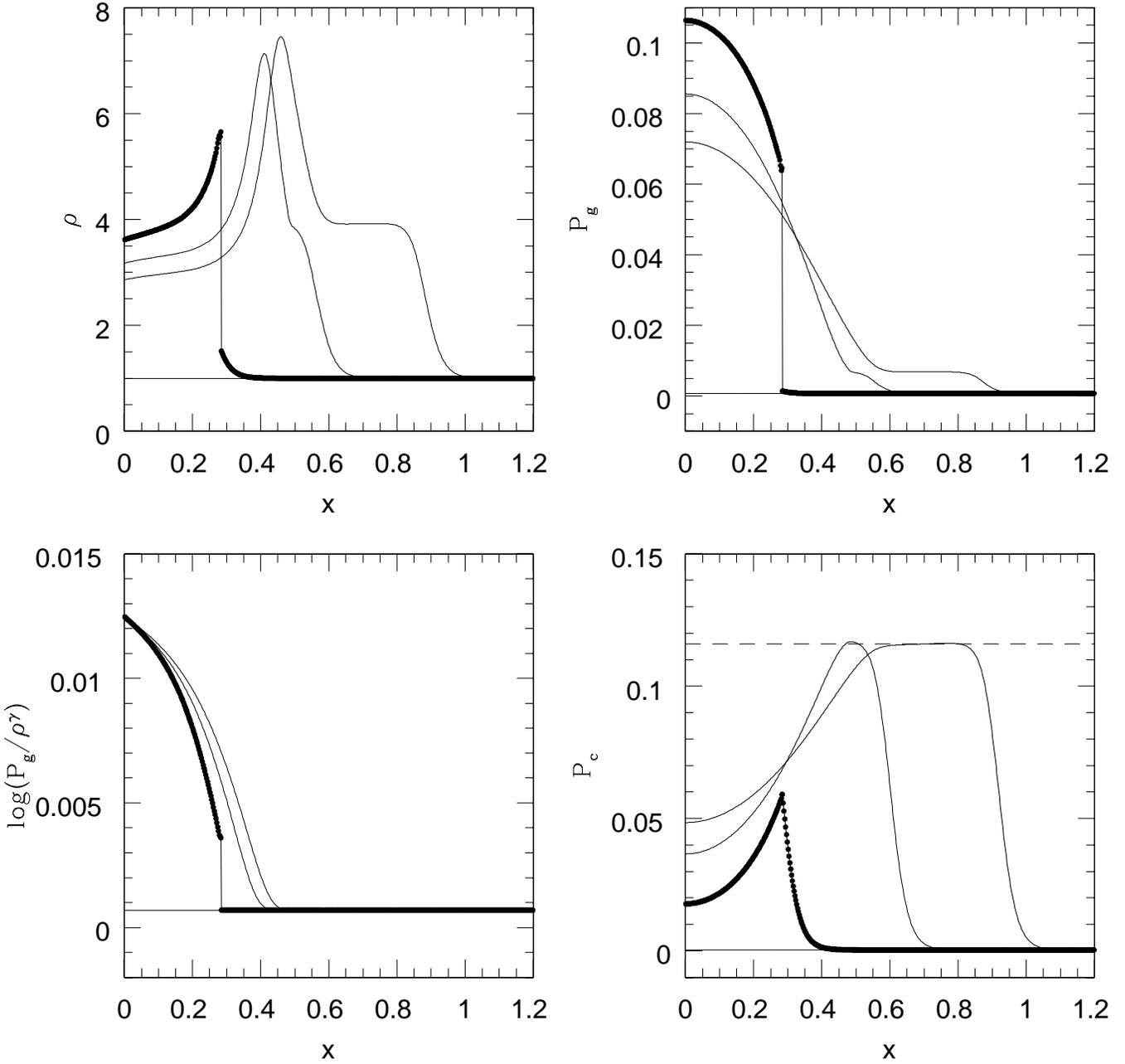}
\figcaption{
A piston-driven shock with $u_p=0.3$, propagating into a uniform medium of
$\rho=1, P_g=7 \times 10^{-4}, P_c=3.5 \times 10^{-4}, \gamma_c=5/3$, and
$<\kappa>=0.01$. CR energy injection at the subshock is not included. 
The simulation was calculated with the two-fluid version of our CRASH code
with the modified entropy ($S$) equation. 
The results are shown at $t=3, 6, $ and 9.
The dashed line corresponds to the postshock $P_c$ in the steady state
limit.
\label{fig13}}
\end{figure}
\clearpage

\begin{figure}
\plotone{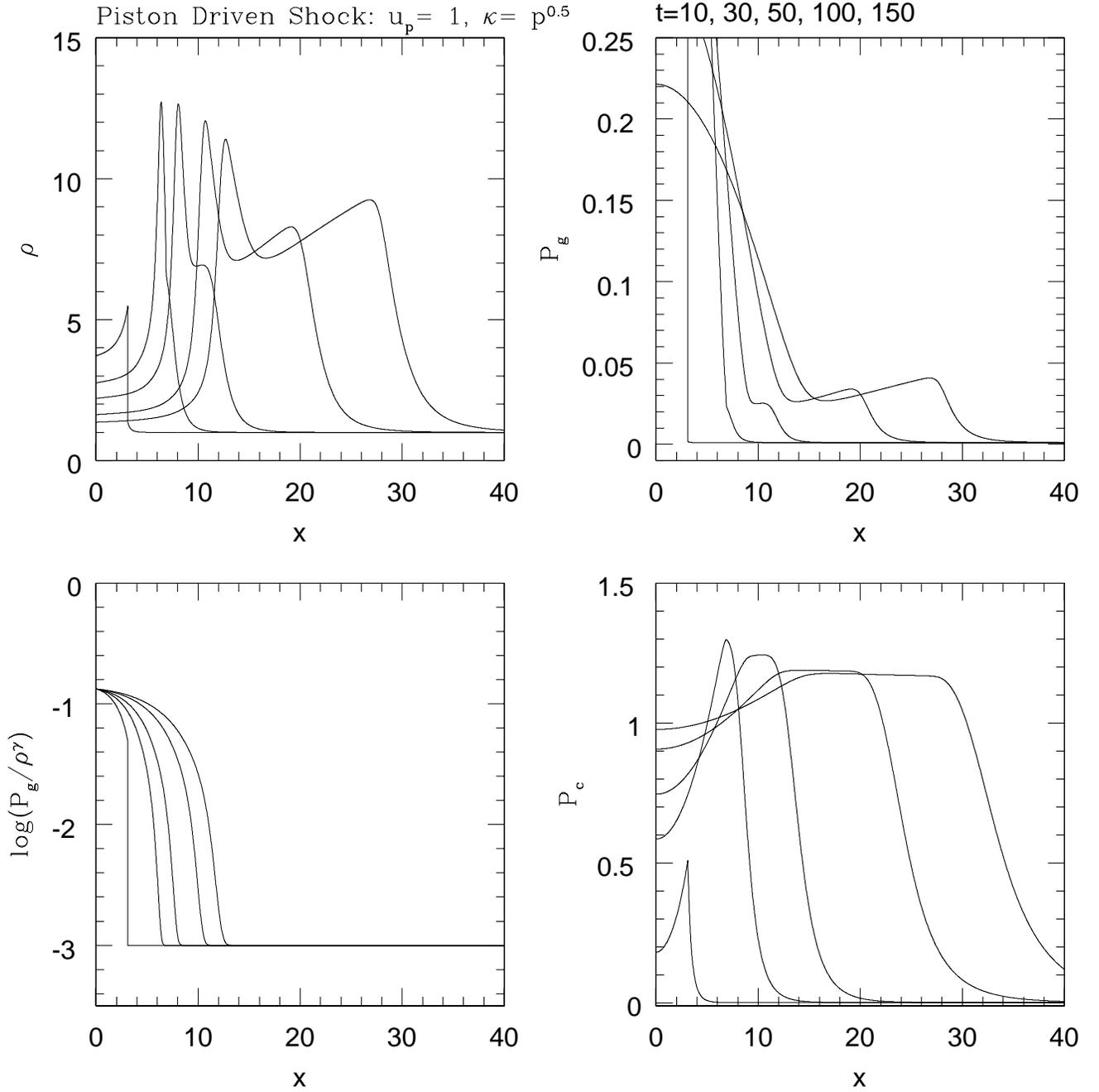}
\figcaption{
A piston-driven shock with $u_p= 1.$, propagating into a uniform medium of
$\rho=1, P_g= P_c=0.001$, and $\kappa=p^{0.5}$. 
CR particle injection at the subshock was turned off. 
The simulation was calculated with the kinetic version of our CRASH code
with the modified entropy ($S$) equation. 
The results are shown at $t=10, 30, 50, 100$ and 150.
\label{fig14}}
\end{figure}
\clearpage

\begin{figure}
\plotone{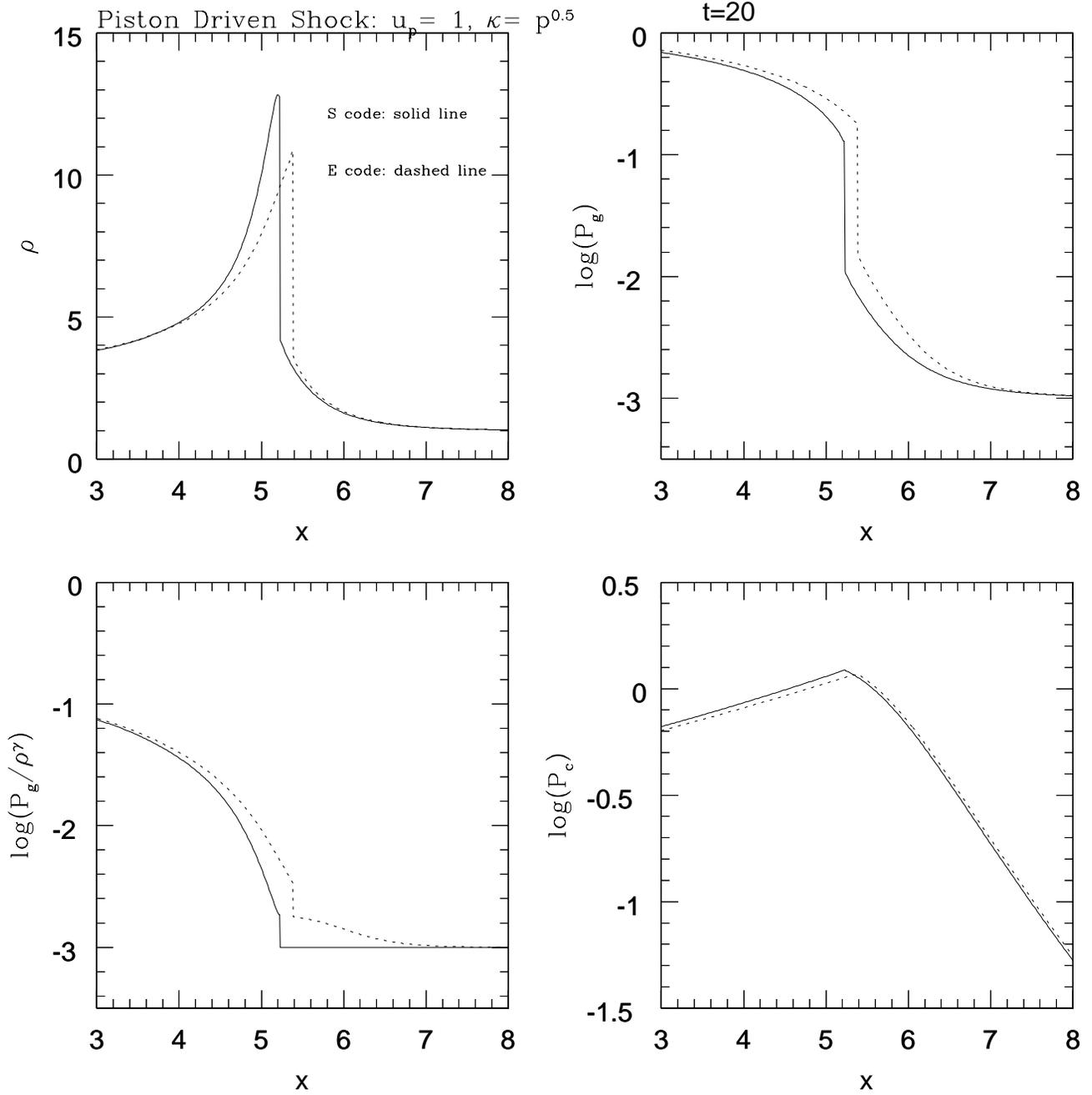}
\figcaption{
The same shock shown in Fig. 14 except at $t=20$. 
The solid lines are for the simulation with the modified entropy
equation, while the dashed lines are for the simulation with the
total energy equation only. 
\label{fig15}}
\end{figure}
\clearpage


\begin{thebibliography}{}

\bibitem[Bell(1978)]{Bell78} Bell A.R., 1978, \mnras, 182, 147

\bibitem[Berezhko and Krymskii(1988)] {berzkry88} Berezhko, E.~G., 
  and Krymskii, G.~F. 1988, Soviet Phys. Usp., 31, 27

\bibitem[Berezhko et al.(1995)]{berz95} Berezhko, E. G., Ksenofontov, 
L., and Yelshin, V. 1995, Nuclear Phys. B, 39A, 171

\bibitem[Berezhko and Ellison(1999)]{berzell99}
Berezhko, E. G. and Ellison, D. C. 1999, \apj, 526, 385 

\bibitem[Berezhko and V\"olk(2000)]{berzvolk00} Berezhko E.G., 
  and V\"olk H.J. 2000, \aap 357, 283

\bibitem[Berger and LeVeque(1998)]{berglev98} Berger, J. S., and LeVeque, 
  R. J. 1998, SIAM J. Numer. Anal., 35, 2298 

\bibitem[Blandford and Eichler(1987)]{blaeic87} Blandford, R.~D., and 
  Eichler, D. 1987, Phys. Rept., 154, 1

\bibitem[Ellison and Eichler(1985)]{elleich85} Ellison, D.~C., and Eichler,
 D. 1985, \apj, 286, 691

\bibitem[Drury(1983)]{dru83} Drury, L.~O'C. 1983, Rept. Prog. Phys., 46, 973

\bibitem[Drury and Falle(1986)]{drufal86} Drury, L.~O'C., and Falle,
  S.~A.~E.~G. 1986, \mnras, 223, 353

\bibitem[Drury, V\"olk and Berezhko(1995)]{dvb95}
Drury, L. O'C., V\"olk, H. J. and Berezhko, E. C. 1995, \aap, 299, 222

\bibitem[Gieseler et al.(2000)]{gies00} Gieseler U.D.J., Jones T.W., 
  and Kang H. 2000, \aap, 364, 911 

\bibitem[Jones and Kang(1990)]{jk90} Jones, T. W. and Kang, H. 1990, \apj, 363, 499

\bibitem[Kang and Jones(1991)]{kanjon91} Kang, H., and Jones, T.~W. 
  1991, \mnras, 249, 439

\bibitem[Kang, Jones and Ryu (1992)]{kjr92} Kang, H., Jones, T.~W., and Ryu, D. 1992, \apj, 385, 193 

\bibitem[Kang and Jones(1995)]{kanjon95} Kang H., and Jones T.W. 1995, 
   \apj, 447, 944 

\bibitem[Kang and Jones(1997)]{kanjon97} Kang H., and Jones T.W. 1997, 
   \apj, 476, 875 

\bibitem[Kang, Jones and Ryu(1992)]{kanjonryu92} Kang, H., Jones, T. W. and 
Ryu, D. 1992, \apj, 385, 193

\bibitem[Kang et al.(2001)]{kang01} Kang, H., Jones, T. W.,
LeVeque, R. J., and Shyue, K. M. ApJ, 550, 737, 2001.

\bibitem[LeVeque and Shyue(1995)]{levshy95} LeVeque, R. J., and 
  Shyue, K. M. 1995, SIAM J. Scien. Comput. 16, 348

\bibitem[Malkov(1997)]{mal97} Malkov M.A. 1997, \apj,  485, 638 

\bibitem[Malkov(1998)]{mal98} Malkov M.A. 1998, Phys. Rev. E,  58, 4911 

\bibitem[Malkov, Diamond and V\"olk(2000)]{mdv00}
Malkov, M. A., Diamond, P. H. and V\"olk, H. J. 2000, apjl, 533, L171

\bibitem[Malkov and V\"olk(1995)]{malvol95} Malkov M.A., and V\"olk H.J. 1995,
\aap, 300, 605

\bibitem[Malkov and V\"olk(1998)]{malvol98} Malkov M.A., and V\"olk H.J. 1998,
Adv. Space Res. 21, 551

\bibitem[Malkov and Drury(2001)]{madru01} Malkov M.A., and Drury, L.O'C. 2001,
Rep. Progr. Phys. 64, 429

\bibitem[Quest(1988)]{Quest88}
Quest K.B., 1988, J. Geophys. Res. 93, 9649

\bibitem[Ryu et~al.(1993)]{rokc93} Ryu, D., Ostriker, J.~P., Kang, H.,
and Cen, R. 1993, \apj, 414, 1

\bibitem[Skilling(1975)]{ski75} Skilling, J. 1975, \mnras, 172, 557

\end{thebibliography}
\end{document}